%% file: DYB_cpc_2012_arx.tex
\documentclass[twoside]{cpc-hepnp}

\usepackage{multicol}
\usepackage{graphicx}
\usepackage{booktabs}
\usepackage{amssymb,bm,mathrsfs,bbm,amscd}
\usepackage[tbtags]{amsmath}
\usepackage{lastpage}

\newcommand{\nuebar}{$\overline{\nu}_{e}$}

\begin{document}

\fancyhead[co]{\footnotesize F.P. An~ et al: Improved Measurement of Electron Antineutrino Disappearance at Daya Bay}
\footnotetext[0]{Received 23 October 2012, revised 15 November 2012}

\title{Improved Measurement of Electron Antineutrino Disappearance at Daya Bay}

\input{dyb_authors_cpc.tex}

%\linenumbers

\begin{abstract}
We report an improved measurement of the neutrino mixing angle $\theta_{13}$ from the Daya Bay Reactor Neutrino Experiment.  We exclude a zero value for $\sin^22\theta_{13}$ with a significance of 7.7 standard deviations.  Electron antineutrinos from six reactors of 2.9 GW$_{\rm th}$ were detected in six antineutrino detectors deployed in two near (flux-weighted baselines of 470 m and 576 m) and one far (1648 m) underground experimental halls.  Using 139 days of data, 28909 (205308) electron antineutrino candidates were detected at the far hall (near halls). The ratio of the observed to the expected number of antineutrinos assuming no oscillations at the far hall is $0.944\pm 0.007({\rm stat.}) \pm 0.003({\rm syst.})$.  An analysis of the relative rates in six detectors finds $\sin^22\theta_{13}=0.089\pm 0.010({\rm stat.})\pm0.005({\rm syst.})$ in a three-neutrino framework.
\end{abstract}

\begin{keyword}
neutrino oscillation, neutrino mixing, reactor, Daya Bay
\end{keyword}

\begin{pacs}
14.60.Pq, 29.40.Mc, 28.50.Hw, 13.15.+g
\end{pacs}

%\footnotetext[0]{\hspace*{-2em}\small\centerline{\thepage\ --- \pageref{LastPage}}}
\footnotetext[0]{\hspace*{-2em}}

\begin{multicols}{2}

\section{Introduction}
\par

Observations of neutrinos and antineutrinos produced in the sun, the atmosphere, reactors, and from particle beams provide overwhelming evidence that the flavors of neutrinos change (oscillate)~\cite{SNO,SuperK,KamLAND,MINOS,K2K}.  The preponderance of data support a three-neutrino framework where three flavor states ($\nu_e, \nu_{\mu}, \nu_{\tau}$) are superpositions of three mass states ($\nu_1, \nu_2, \nu_3$).  This mixing can be quantified using a unitary $3\times 3$ mixing matrix described in terms of three mixing angles ($\theta_{12},\theta_{23},\theta_{13}$) and a CP violating phase ($\delta$)~\cite{pontecorvo,mns}.    Neutrino oscillations are also dependent on the differences in the squares of the neutrino masses.

The Daya Bay collaboration recently measured a non-zero value for $\sin^2 2\theta_{13}=0.092\pm 0.016({\rm stat.})\pm0.005({\rm syst.})$~\cite{DB_discovery}, an observation consistent with previous and subsequent experimental results~\cite{RENO,DC,T2K,MINOS}.  In absolute terms, the value of $\theta_{13}$ is now known with better precision than either of the other two mixing angles.  Constraining the value of $\theta_{13}$ increases the constraints on the other mixing parameters (mixing angles and mass squared differences) through a global fit of all available oscillation data~\cite{Fogli,Tortola}.

For reactor-based experiments, in a three-neutrino framework, an unambiguous determination of $\theta_{13}$ can be extracted via the survival probability of the electron antineutrino \nuebar\ at short distances (${\cal O}(km)$) from the reactors
\begin{equation}\label{eqn:psurv}
P_{\rm sur} \approx 1 - \sin^2 2\theta_{13} \sin^2 (1.267 \Delta m^2_{31}  L/E) \,,
\end{equation}
where $\Delta m^2_{31}$ can be approximated by $\Delta m^2_{atm}=(2.32^{+0.12}_{-0.08}) \times 10^{-3} {\rm eV}^2$~\cite{minosdm}, $E$ is the \nuebar\ energy in MeV and $L$ is the distance in meters between the \nuebar\ source and the detector (baseline).  The near-far arrangement of antineutrino detectors (ADs), as illustrated in Fig.~\ref{fig:layout}, allows for a relative measurement by comparing the observed \nuebar\ rates at various distances.  With functionally identical ADs, the relative rate is independent of correlated uncertainties, and uncorrelated reactor uncertainties are minimized.

\par
The results reported here were derived using the same analysis techniques and event selection as our previous results~\cite{DB_discovery}, but were based on data collected between December 24, 2011 and May 11, 2012,  a 2.5 fold increase in statistics.
A blind analysis strategy was adopted for our previous results, with the baselines, the thermal
power histories of the cores, and the target masses of the ADs hidden until the analyses were finalized. Since the baselines and the target masses have been unveiled for the six ADs, we kept the thermal power histories hidden in this analysis until the analyses were finalized.

%\begin{figure}[htb]
\begin{center}
\includegraphics[width=8cm]{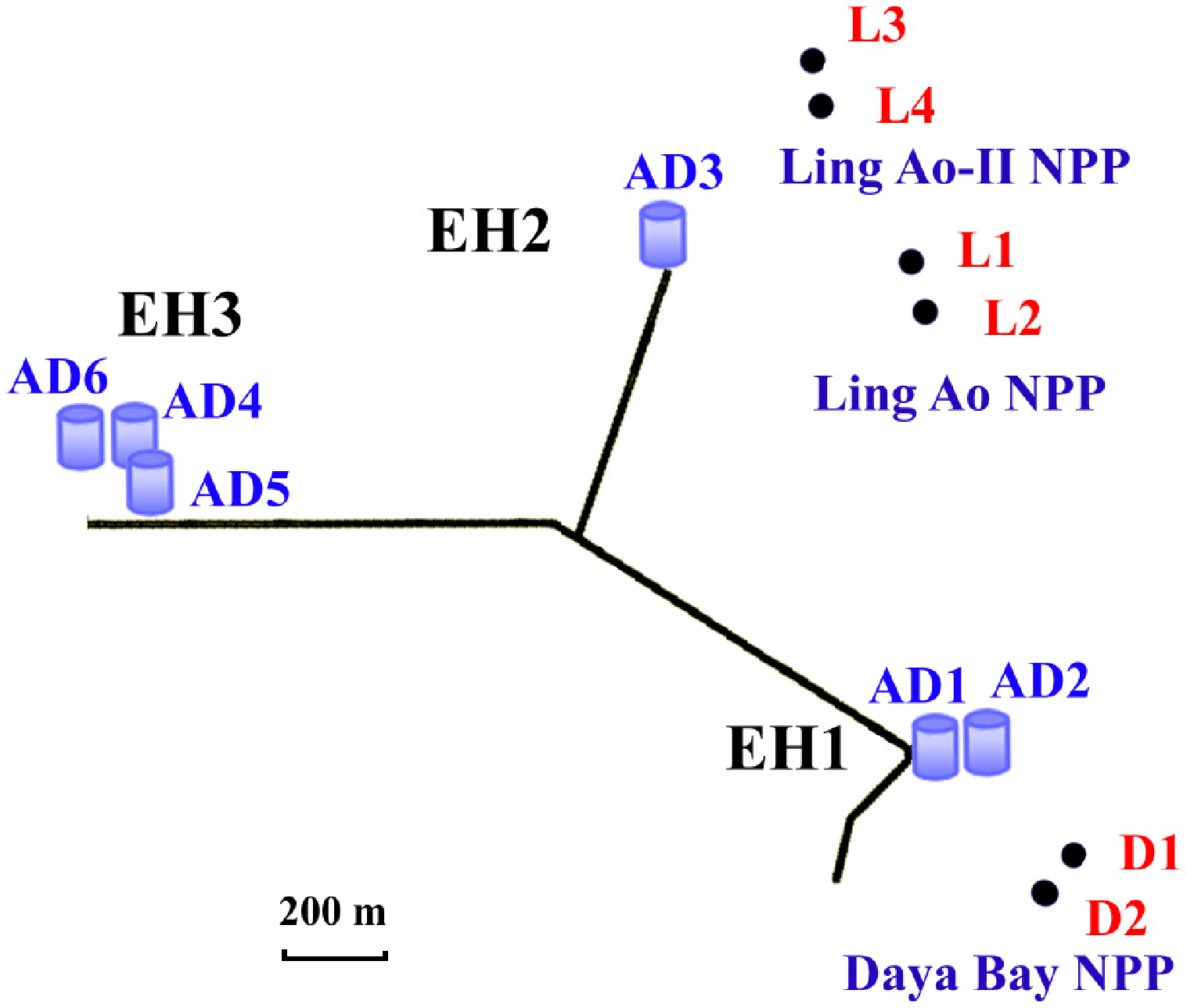}
\figcaption{Layout of the Daya Bay experiment.
The dots represent reactor cores, labeled as D1, D2, L1, L2, L3 and L4.
Six antineutrino detectors (ADs) were installed in three experimental halls (EHs). \label{fig:layout}}
\end{center}
%\end{figure}

\section{The Experiment}
\subsection{Site}

The Daya Bay nuclear power complex is located on the southern coast of China, 55 km to the northeast of Hong Kong and 45 km to the east of Shenzhen. A detailed description of the Daya Bay experiment can be found in~\cite{dyb,ad12}.  As shown in Fig.~1, the nuclear complex consists of six reactors grouped into three pairs with each pair referred to as a nuclear power plant (NPP)\@.   All six cores are functionally identical pressurized water reactors, each with a maximum of 2.9 GW thermal power~\cite{Guangdong}. The last core started commercial operation on Aug.\ 7, 2011. The distance between the cores for each pair is 88 m. The Daya Bay cores are separated from the Ling Ao cores by about 1100 m, while the Ling Ao-II cores are around 500 m away from the Ling Ao cores.

%\begin{table}[htb]
\begin{center}
\tabcaption{Vertical overburden, muon rate $R_\mu$, and average muon energy $<E_\mu>$  of the three EHs. \label{tab:murate}}
\footnotesize
\begin{tabular}[c]{cccc}
\toprule
& Overburden (m.w.e) & $R_\mu$ (Hz/m$^2$) & $<E_\mu>$ (GeV) \\\hline
EH1 & 250 & 1.27 & 57 \\
EH2 & 265 & 0.95 & 58 \\
EH3 & 860 & 0.056 & 137 \\
\bottomrule
\end{tabular}
\end{center}
%\end{table}

\par
Three underground experimental halls (EHs) are connected with horizontal tunnels. For this analysis, two antineutrino detectors (ADs) were located in EH1, one in EH2, and three near the oscillation maximum in EH3 (the far hall). The overburden in equivalent meters of water (m.w.e.), simulated muon rate and average muon energy are listed in Table~\ref{tab:murate}.

\par
The distances from the six ADs to the six cores are listed in Table~\ref{tab:baseline}.  All distances have been surveyed with the Global Positioning System (GPS) and with modern theodolites utilizing two major control networks built over several months.  The network surveyed using GPS is within the campus of the power plant but outside of the tunnel. The other network is inside the tunnel system, surveyed using Total Station, an electronic/optical instrument widely used in modern surveying.  The double traverse survey network was laid down in a closed ring in the 7-m wide tunnels.  The Total Station survey included the power plant campus to link the two control networks. The survey from the anchors at the entrance of each experimental hall to each AD was completed during the installation of each AD using a laser tracker. The coordinates of the AD center were further deduced using the AD survey data collected during AD assembly. The coordinates of the geometrical center of the reactor cores were provided by the power plant relative to four anchor points outside of each nuclear island. The survey data were processed independently by three groups with different software. The uncertainty of the baselines was determined to be 28 mm as reported in Ref.~\cite{DB_discovery}. Recently another closed traverse survey was completed utilizing a different tunnel entrance and the top of the mountain. The largest baseline difference between the two surveys is 4 mm and the uncertainty in the baselines has been reduced to 18 mm. The uncertainty has seven significant contributions, the largest being 12.6 mm due to the precision of the GPS survey.  The second largest is 9.1 mm due to fitting uncertainties associated with the linking of the GPS and the Total Station networks. When combined with the uncertainties of the fission gravity center (described in Sec.~6), the baseline uncertainties were found to make a negligible contribution to the oscillation uncertainties.

%\begin{table}[htb]
\begin{center}
\tabcaption{Baselines from antineutrino detectors AD1-6 to reactors D1, D2, and L1-4 in meters.\label{tab:baseline}}
\footnotesize
\begin{tabular}[c]{cccrrrr}
\toprule
& D1 & D2 & L1 & L2 & L3 & L4  \\
\hline
AD1 & 362 & 372 & 903 & 817 & 1354 & 1265 \\
AD2 & 358 & 368 & 903 & 817 & 1354 & 1266 \\
AD3 & 1332 & 1358 & 468 & 490 & 558 & 499 \\
AD4 & 1920 & 1894 & 1533 & 1534 & 1551 & 1525 \\
AD5 & 1918 & 1892 & 1535 & 1535 & 1555 & 1528 \\
AD6 & 1925 & 1900 & 1539 & 1539 & 1556 & 1530 \\
\bottomrule
\end{tabular}
\end{center}
%\end{table}
\vspace{0.2cm}

\subsection{Antineutrino Detectors}

\par

The $\overline{\nu}_{e}$s are detected via the inverse $\beta$-decay (IBD) reaction, $\overline{\nu}_e + p \to e^+ + n$, in gadolinium-doped liquid scintillator (Gd-LS)~\cite{ding,yeh}\@. The coincidence of the prompt scintillation from the $e^+$ and the delayed neutron capture on Gd provides a distinctive \nuebar\ signature. The positron carries almost all of the kinetic energy of the antineutrino, thus the positron energy deposited in the liquid scintillator is highly correlated with the antineutrino energy. The neutron thermalizes before being captured on either a proton or a gadolinium nucleus with a mean capture time of $\sim$30 $\mu$s in Gd-LS with 0.1\% Gd by weight. When a neutron is captured on Gd, it releases several gamma-rays with a total energy of $\sim$8~MeV, and is thus easily distinguished from the background coming from natural radioactivity. Only neutrons that captured on Gd were selected as the delayed signal of a antineutrino event in this analysis.

\par
Each AD has three nested cylindrical volumes separated by concentric acrylic vessels~\cite{AVPaper} as shown in Fig.~\ref{fig:det}.  The innermost volume holds 20 t of Gd-LS with 0.1\% Gd by weight and serves as the antineutrino target. The middle volume is called the gamma catcher and is filled with 20 t of un-doped liquid scintillator (LS) for detecting gamma-rays that escape the target volume.  The outer volume contains 37 t of mineral oil (MO) to provide optical homogeneity and to shield the inner volumes from radiation originating, for example, from the photo-multiplier tubes (PMTs) or the stainless steel vessel (SSV).  There are 192 20-cm PMTs (Hamamatsu R5912) installed along the circumference of the SSV and within the mineral oil volume, in 24 columns and 8 rings. To improve optical uniformity, the PMTs are recessed in a 3-mm thick black acrylic cylindrical shield located at the equator of the PMT bulb.

%\begin{figure}[htb]
\begin{center}
\includegraphics[width=\columnwidth]{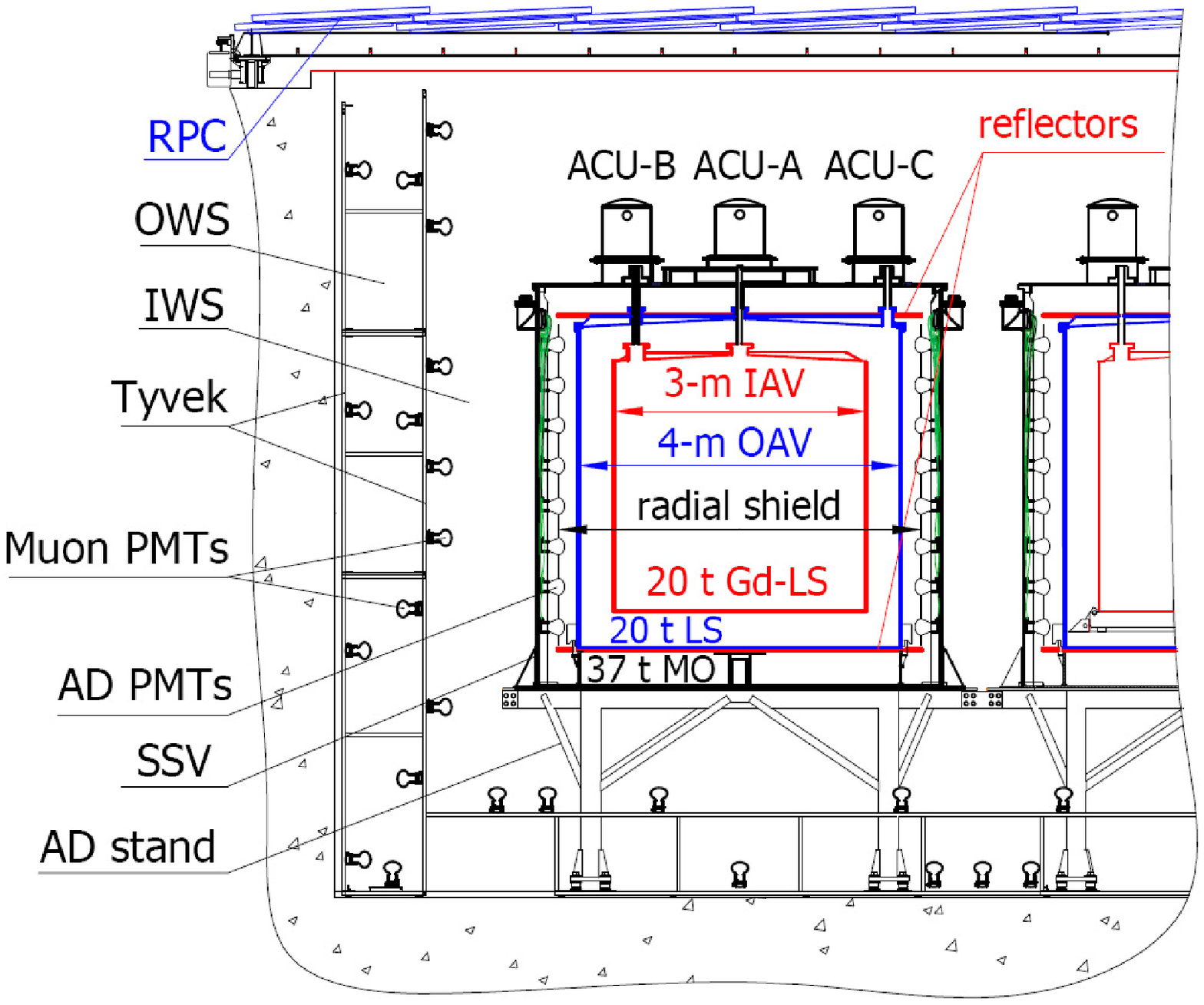}
\figcaption{Schematic diagram of the Daya Bay detectors. \label{fig:det}}
\end{center}

\par
Three automated calibration units (ACU-A, ACU-B, and ACU-C) are mounted on the top of each SSV as shown in Fig.~\ref{fig:det}. Each ACU is equipped with a LED, a $^{68}$Ge source, and a combined source of $^{241}$Am-$^{13}$C and $^{60}$Co. The Am-C source generates neutrons at a rate of $\sim$0.5 Hz. The rates of the $^{60}$Co and $^{68}$Ge sources are about 100 Hz and 15 Hz, respectively.  Since the AD is fully submerged in water, the ACUs are operated remotely.  The sources can be deployed to better than 0.5 cm along a vertical line down to the bottom of the acrylic vessels.  When not in use, the LED and sources are retracted into the ACUs that also serve as shielding for the sources.

\subsection{Muon System}
\par

The muon detection system consists of a resistive plate chamber (RPC) tracker and a high-purity active water shield.  The water shield consists of two optically separated regions known as the inner (IWS) and outer (OWS) water shields. There are 121 (160) PMTs installed in the IWS and 167 (224) PMTs in the OWS in each near (far) hall. Each region operates as an independent water Cherenkov detector.  The muon detection efficiency is 99.7\% and 97\% for the IWS and OWS, respectively~\cite{ad12}. In addition to detecting muons that can produce spallation neutrons or other cosmogenic backgrounds in the ADs, the pool moderates neutrons and attenuates gamma rays produced in the rock or other structural materials in and around the experimental hall.  At least 2.5 m of water surrounds the ADs in every direction.  Each pool is outfitted with a light-tight cover overlaying a dry-nitrogen atmosphere.

Each water pool is covered with an array of RPC modules~\cite{Zhang:2007ap,RPC_IEEE}.  The 2~m $\times$ 2~m modules are layered on a steel frame to minimize dead areas.  The assembly is mounted on rails and can be retracted to provide access to the water pool.  There are four layers of bare RPCs inside each module, with one layer of readout strips associated with each layer of bare RPCs.  The strips have a ``switchback" design with an effective width of 25 cm, and are stacked in alternating orientations providing a spatial resolution of $\sim$8 cm.

\subsection{Trigger and Readout}

Each detector unit (AD, IWS, OWS, and RPC) is read out with a separate VME crate.   All PMT readout crates are physically identical, differing only in the number of instrumented readout channels.  The front-end electronics board (FEE) receives raw signals from up to sixteen PMTs, sums the charge from all input channels, identifies over-threshold channels, records their timing information, and measures the charge of each over-threshold pulse with a 40 MHz sampling rate~\cite{FEE_IEEE}. The FEE in turn sends the number of channels over threshold and the integrated charge to the trigger system.  When a trigger is issued, the FEE reads out the charge and timing information within 1 $\mu$s for each over-threshold channel, as well as the average ADC value over a 100 ns time-window immediately preceding the over-threshold condition (preADC).

Triggers are primarily created internally within each PMT readout crate based on the number of over-threshold channels (NHIT) as well as the summed charge (E-Sum) from each FEE~\cite{trigger_nim}.  The system is also capable of accepting external trigger requests, for example, from the calibration system.  The trigger system blocks triggers when either the trigger data-buffer or a FEE data-buffer is nearly full.  The number of blocked triggers is recorded and read out for calculating the dead time offline.

\section{Data Characteristics, Calibration and Modelling}
\subsection{Data set}

The data used in this analysis were collected from December~24,~2011 through May~11,~2012.  Table~\ref{tbl:data} summarizes the experimental livetime for each hall. Total data acquisition (DAQ) time measures the number of hours that the DAQ was collecting data, with about 2\% of the DAQ time devoted to detector calibration. Standard data running (Physics Data or Physics DAQ time) accounted for more than 93\% of the calendar time.  We further rejected about 60 hours of physics data from each hall due to excessive coherent electromagnetic pickup, PMT high voltage (HV) trips, electronic or DAQ problems, or requirements of simultaneous operation in all three halls. The resulting data set (Good run data or Good run time) were used for analysis.

%\begin{table}[!ht]
\begin{center}
\tabcaption{Summary of experimental livetime in hours.
\label{tbl:data}}
\footnotesize
\begin{tabular}{ l l l l }
\toprule
& EH1 & EH2 & EH3 \\\hline
 Total calendar time  & 3322.1 & 3322.1 & 3322.1\\
 Total DAQ time & 3195.4 & 3179.5 & 3171.6 \\
 Physics DAQ time  & 3117.9 & 3122.0 & 3093.6\\
 Good run time  & 3061.1 & 3057.1 & 3030.5 \\
%\hline\hline
\bottomrule
\end{tabular}
\end{center}
%\end{table}

\par
The detector halls operated independently with a common centralized clock and GPS timing system.  The analysis presented here required simultaneous operation of all three detector halls, to minimize systematic effects associated with potential reactor power excursions. Simultaneous operation was defined as Physics Data within a given hour existing for all three detector halls. The data samples used in this analysis differed by 1\% in time for the three halls. A more rigorous requirement that demands synchronization among the three halls on the scale of seconds was tested with no change to the reported results.

\subsection{Triggered Detector Rates}

Triggers were formed based either on the number of PMTs with signals above a $\sim$0.25 photoelectron (p.e.) threshold (NHIT triggers), or the charge sum of the PMTs (E-Sum triggers).  AD triggers with NHIT $>$ 45 or E-Sum $\gtrsim$ 65 p.e.\ correspond to an event energy threshold of $\sim$0.4 MeV~\cite{ad12}. The corresponding trigger rate per AD was $<$ 280 Hz with a negligible trigger inefficiency for IBD candidates.

The \nuebar\ candidates were selected in the offline analysis using the coincidence of a prompt signal from the $e^+$ and a delayed signal due to neutron capture on Gd. A prompt-type (delayed-type) signal was defined as an event with energy in the range of 0.7-12 MeV (6-12 MeV). The rates of prompt-type and delayed-type singles that are separated in time by at least 200~$\mu$s from any additional signals with an energy $>$ 0.7 MeV were of particular interest for background studies and detector stability monitoring. They are shown in Fig.~\ref{fig:s3_singles}.  A veto was applied to reject events within -2 to 200~$\mu$s relative to a muon (defined in Sec.~4.1).  The data were corrected for the corresponding inefficiencies. These rates were used to estimate the accidental background rate as described in Sec.~5.1.

%\begin{figure}[htb]
\begin{center}
\includegraphics[width=\columnwidth]{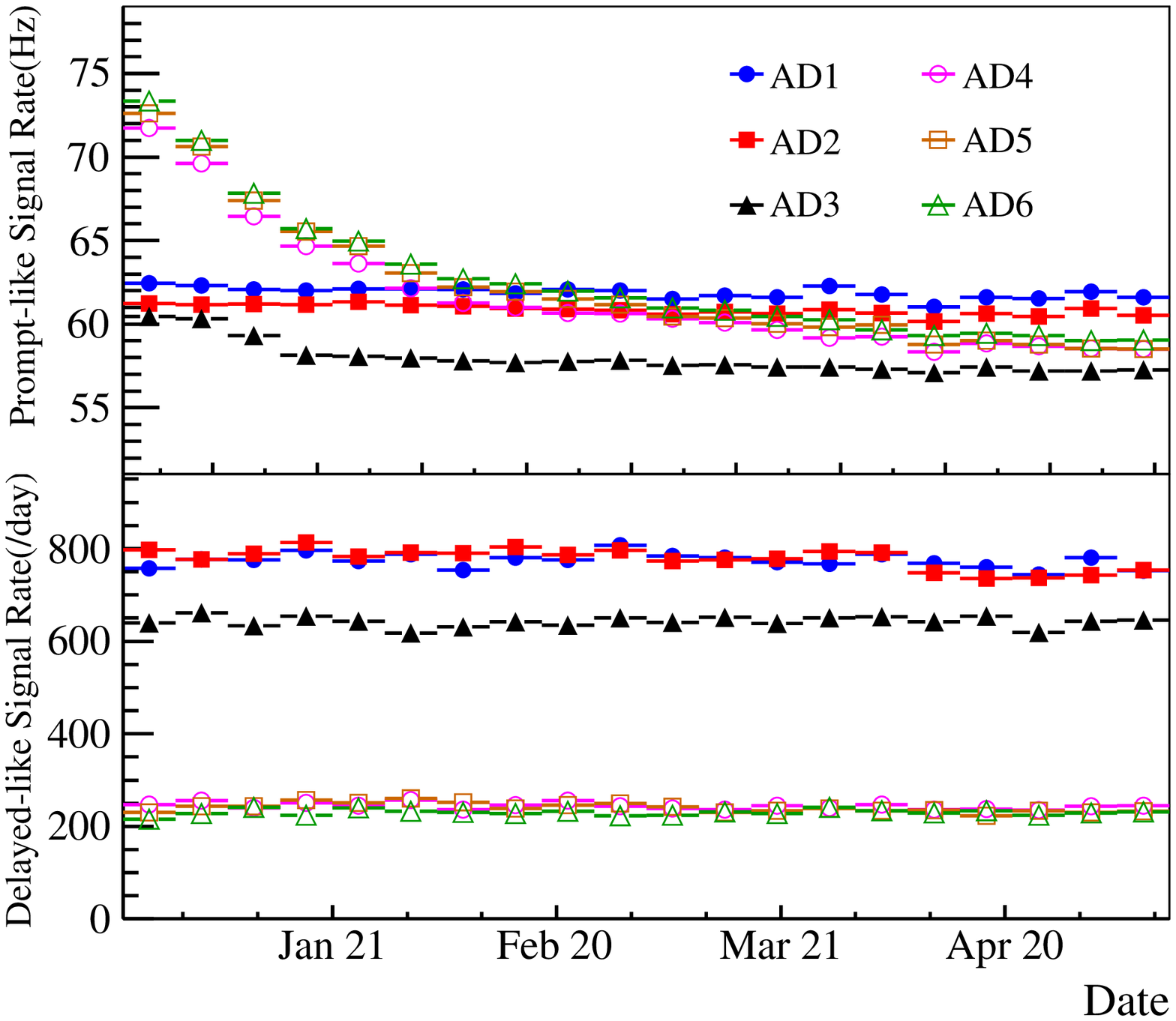}
\figcaption{Singles rates for the six ADs. The top panel shows the prompt candidates and the bottom panel shows the delayed candidates.\label{fig:s3_singles}}
\end{center}
%\end{figure}

The observed rate of low energy signals decreased with time.  The detectors in EH1 initiated data taking on Aug.~15, 2011 and the AD in EH2 started on Nov.~5, 2011.  As such, these detectors (AD1-3) had reached a steady state by December~24, 2011, while the rates in AD4-6 in EH3 exhibited decaying behavior, as shown in Fig.~\ref{fig:s3_singles}.

The muon rates in the water Cherenkov detectors (IWS and OWS) were closely monitored, as shown in Fig.~\ref{fig:s3_WSTrigger}. IWS and OWS events were selected with NHIT $>$ 12. The event rates were different for the three halls due to differing muon rates in each hall and different sizes of the far hall and the near halls.

%\begin{figure}[htb]
\begin{center}
\includegraphics[width=\columnwidth]{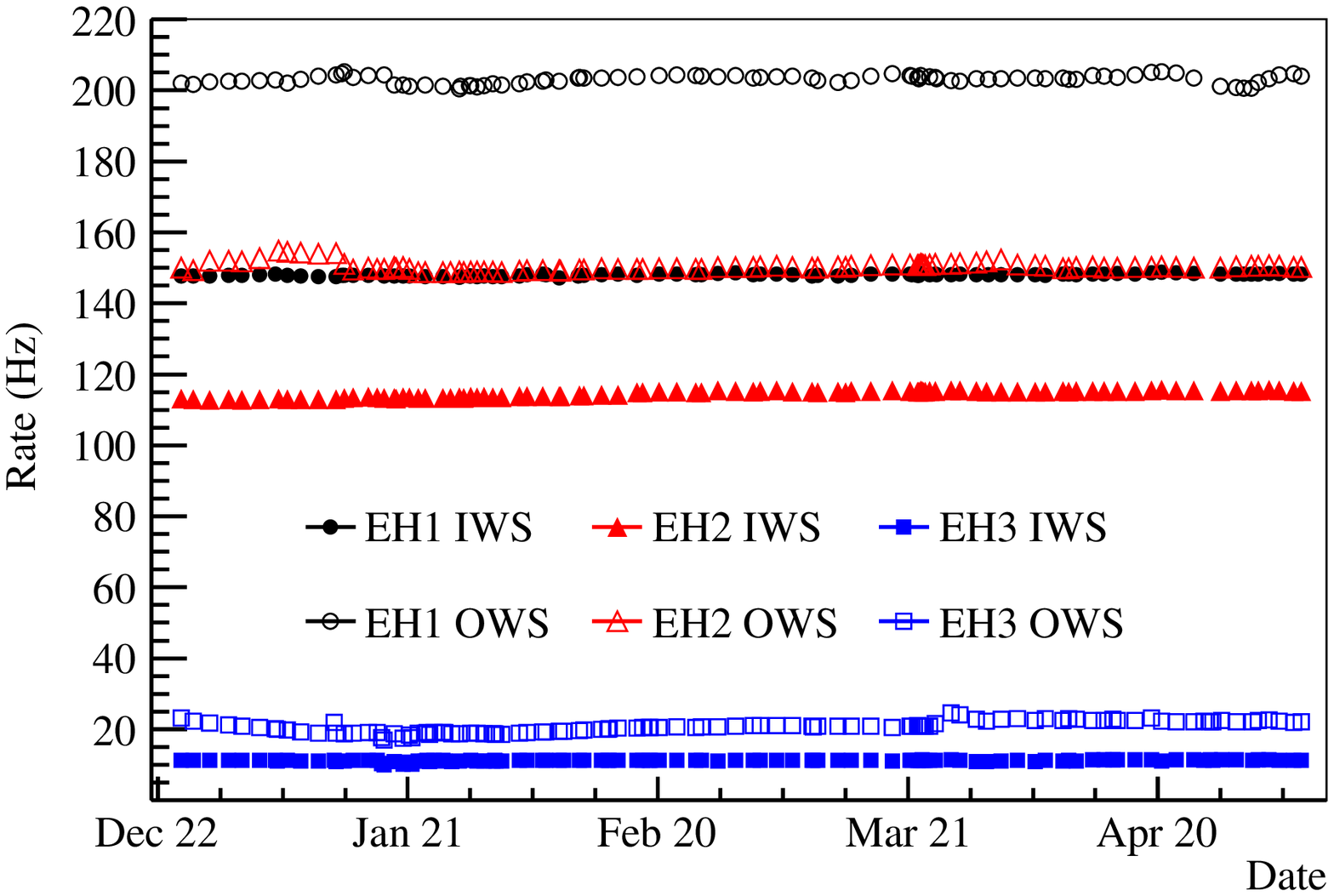}
\figcaption{Muon rates in the inner (IWS) and outer water shield (OWS) in the three experimental halls.
\label{fig:s3_WSTrigger}}
\end{center}
%\end{figure}

\subsection{Instrumental Backgrounds}

A small number of AD PMTs spontaneously emit light, due to discharge within the base.  These instrumental backgrounds are referred to as flasher events.  For Daya Bay, the reconstructed energy of such events covers a wide range, from sub-MeV to 100 MeV. Two features were typically observed when a PMT flashed: the observed charge fraction for a given PMT was very high, and PMTs on the opposite side of the AD saw large fraction of light from the flashed PMT.  The charge pattern of a typical flasher event is shown in Fig.~\ref{fig:s3_flasher_pattern}.

\par
To reject flasher events, two variables, named $MaxQ$ and $Quad$, were created based on the distinctive charge pattern. $MaxQ$ is the largest fraction of the total detected charge seen by a single PMT (the ``hottest" PMT).  There are twenty-four columns of PMTs in an AD that can be divided into four quadrants.  With the hottest PMT centered in the first quadrant,  $Quad$ was defined as $Q_3/(Q_2+Q_4)$, where $Q_i$ is the charge sum of the PMTs in the $i$-th quadrant. A flasher event identification variable (FID) was constructed based on $MaxQ$ and $Quad$:

\begin{equation}
FID=\log_{10}[(MaxQ/0.45)^2 + (Quad)^{2}].
\end{equation}

\noindent Fig.~\ref{fig:s3_flasher_discr} shows the discrimination of flasher events for the delayed signal of the IBD candidates. The distributions for all six ADs agree well for IBD candidates (FID$<$0); however, there is some variation for flasher candidates (FID$>$0). For the IBD analysis as well as other analyses, the rejection of flasher events was done at the beginning of the data reduction.

%\begin{figure}[htb]
\begin{center}
\includegraphics[width=\columnwidth]{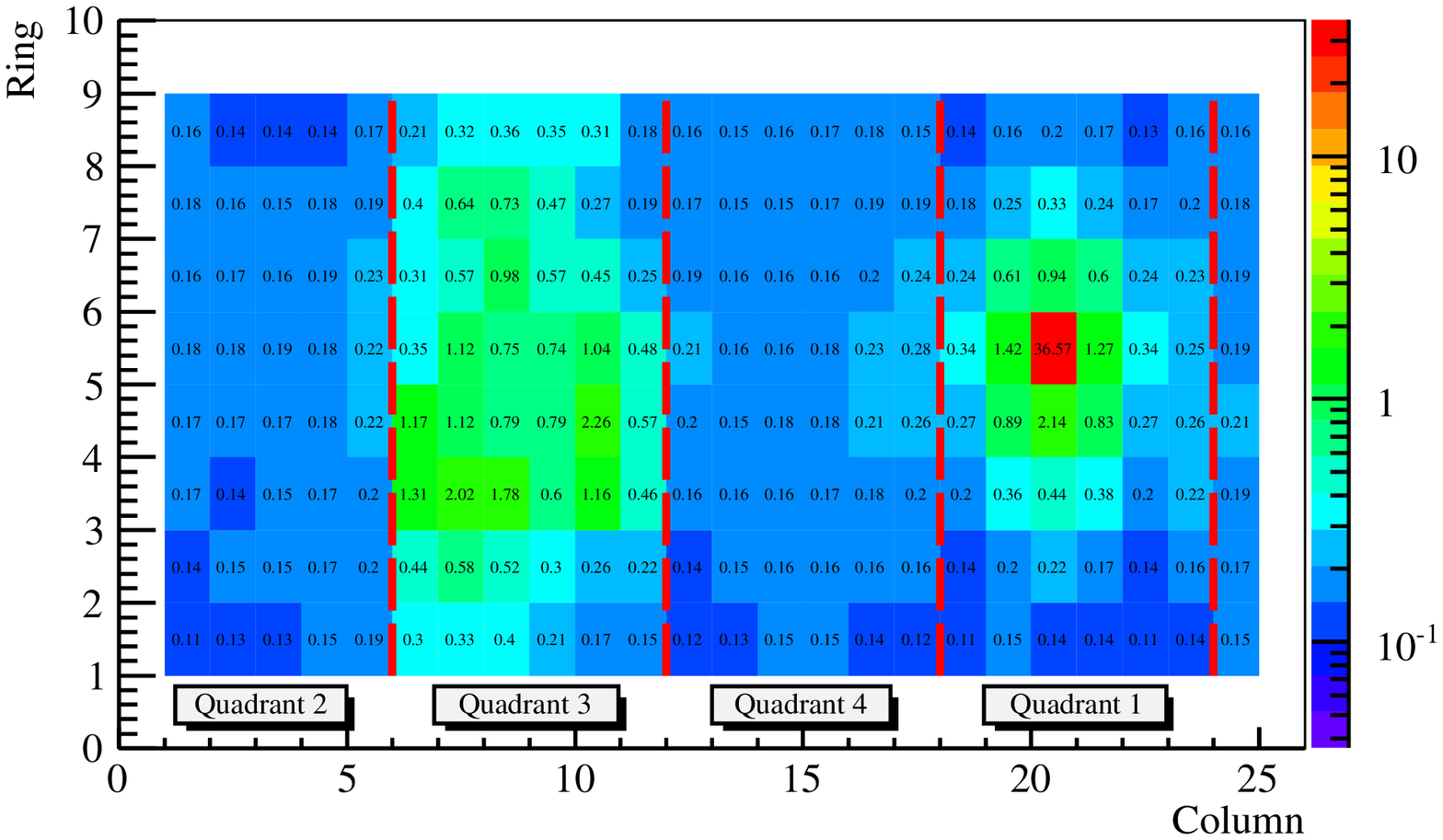}
\figcaption{Topology of a typical flasher event.  Such events are distinctive, characterized by a single channel with substantially more charge than in surrounding PMTs, as well as excessive charge on the opposite side of the AD. \label{fig:s3_flasher_pattern}}
\end{center}
%\end{figure}

%\begin{figure}[htb]
\begin{center}
\includegraphics[width=\columnwidth]{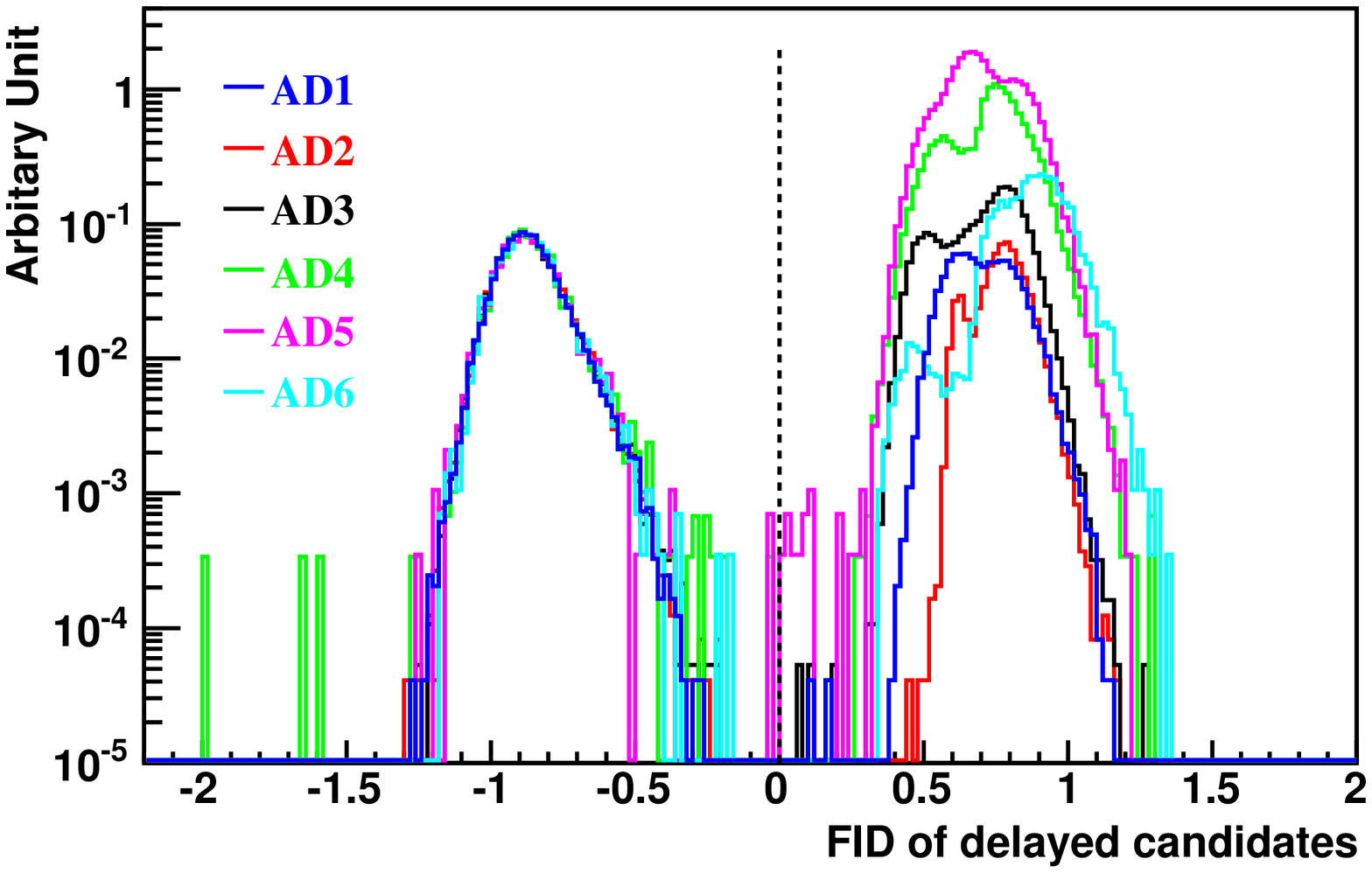}
\figcaption{Discrimination of flasher events (FID $>0$) and IBD delayed signals (FID $<0$). The delayed signals of IBDs have the same distribution for all six ADs while the flashers are different. The FID $<0$ distributions have been scaled to equal area. \label{fig:s3_flasher_discr}}
\end{center}
%\end{figure}

\par
The discrimination power decreases for low energy events or events very close to PMTs. For the rejected events with FID $\sim 0$, we studied the charge pattern, the energy distribution, the capture time, and the distance between the vertices of the prompt and the delayed signals, and found that some were consistent with real IBD events. By counting such events, the inefficiency of the IBD selection due to the flasher rejection was estimated to be 0.02\% with an uncorrelated uncertainty of 0.01\%. The background contamination of selected IBD candidates was evaluated to be $<10^{-4}$. Furthermore, such contamination was counted as accidental background (Sec.~5.1) and was subtracted. Special runs were conducted with reduced high voltage for selected PMTs to cross-check the identified PMTs that exhibited flashing. Due to the high efficiency of the FID, all AD PMTs were kept in operation, including those identified as flashing PMTs.

\subsection{Energy Reconstruction}
In general, the energy response of the AD can depend on time, position in the fiducial volume (non-uniformity), particle species, and their energies (non-linearity). The goal of energy reconstruction was to correct for these dependencies in order to minimize the uncertainties in the AD energy scale.  To achieve this goal, each AD was calibrated using LEDs, $^{68}$Ge, and $^{241}$Am-$^{13}$C/$^{60}$Co sources. LEDs were utilized for PMT gain calibration, while the energy calibration parameter (p.e.\ per MeV) was determined with a $^{60}$Co source deployed at the detector center.  The sources were deployed once per week to check and correct for any time dependence. Occasionally a PMT's output was noisy and was temporarily turned off during physics data taking. The energy calibration corrected such situations. The energy calibration parameter for each AD is shown in Fig.~\ref{fig:s3_ecal_constant} as a function of time. The small jumps correspond to the temporary turn-off of noisy PMTs. The energy resolution was (7.5/$\sqrt{E({\rm MeV})}+0.9)\%$ for all 6 ADs.

%\begin{figure}[htb]
\begin{center}
\includegraphics[width=\columnwidth]{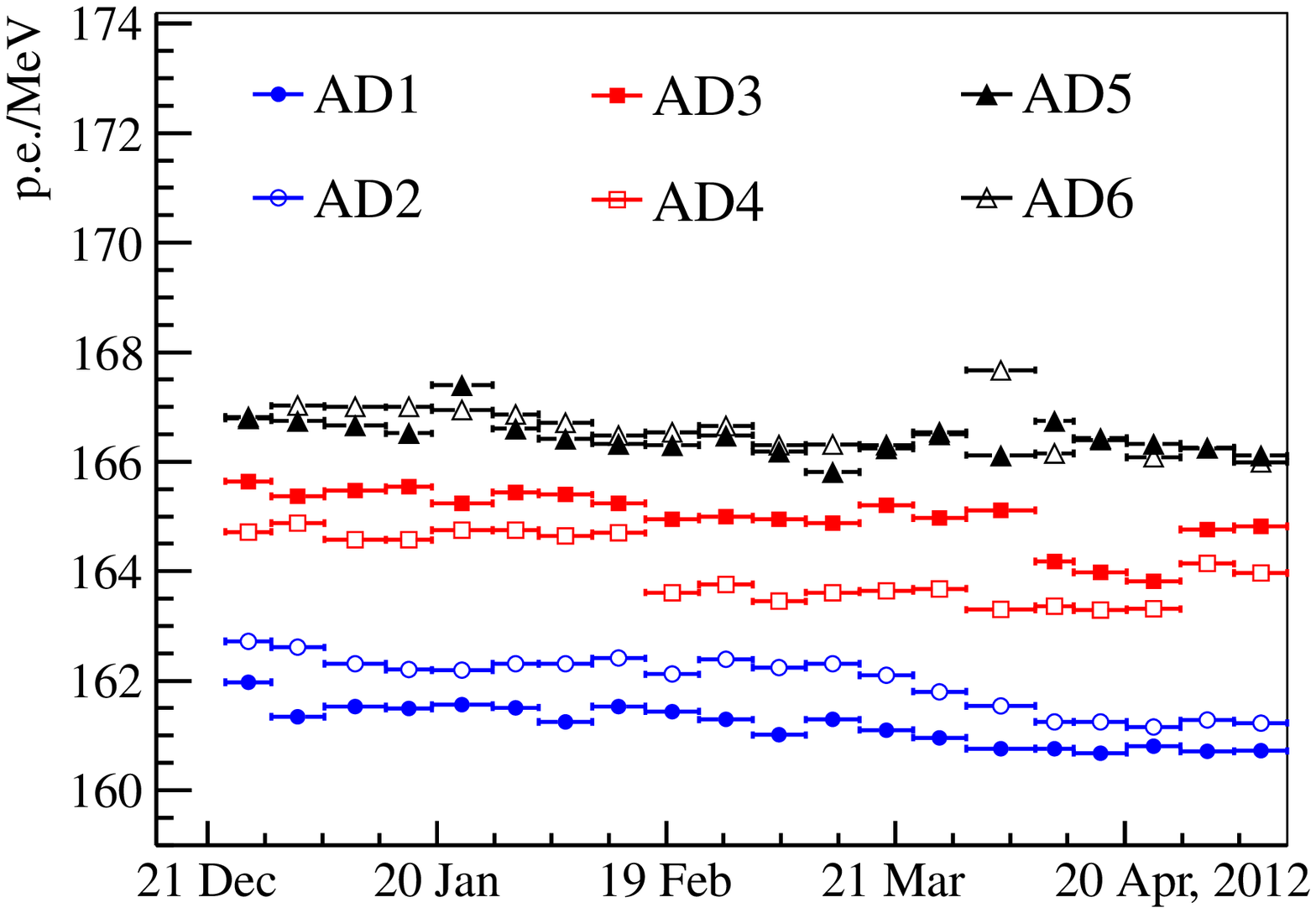}
\figcaption{Calibration parameter versus time for each AD.\label{fig:s3_ecal_constant}}
\end{center}
%\end{figure}

A scan along the vertical axis utilizing the $^{60}$Co source from each of the three ACUs was used to obtain non-uniformity correction functions. The non-uniformity was also studied with spallation neutrons generated by cosmic muons, and alphas produced by natural radioactivity present in the liquid scintillator. The neutron energy scale was set by comparing $^{60}$Co events with neutron capture on Gd events from the Am-C source at the detector center.  Additional details of energy calibration, reconstruction, and vertex reconstruction can be found in Ref.~\cite{ad12}.

The AD energy scale uncertainty was studied by comparing the energy peaks in all ADs using neutron capture on gadolinium from IBD and muon spallation products, alphas from Polonium decay in the Gd-LS, and each of the calibration sources. Asymmetries of the six ADs' response are shown in Fig.~\ref{fig:s3_energy_asym}. For each type of event, we defined the Asymmetry as
\begin{equation}
Asymmetry_i = \frac{E_i-\sum E_i/6}{\sum E_i/6}\,,
\end{equation}
where $E_i$ is the fitted mean energy of the studied type of event of the $i$-th AD.
The energy scale uncertainty was set at 0.5\% in Ref.~\cite{ad12} based on extensive side-by-side studies of AD1 and AD2.  Extending this to six ADs, Asymmetries for all types of events in all the ADs fall within a band of 0.5\%. As such, we kept the same uncertainty, uncorrelated among ADs.

%\begin{figure}[htb]
\begin{center}
\includegraphics[width=\columnwidth]{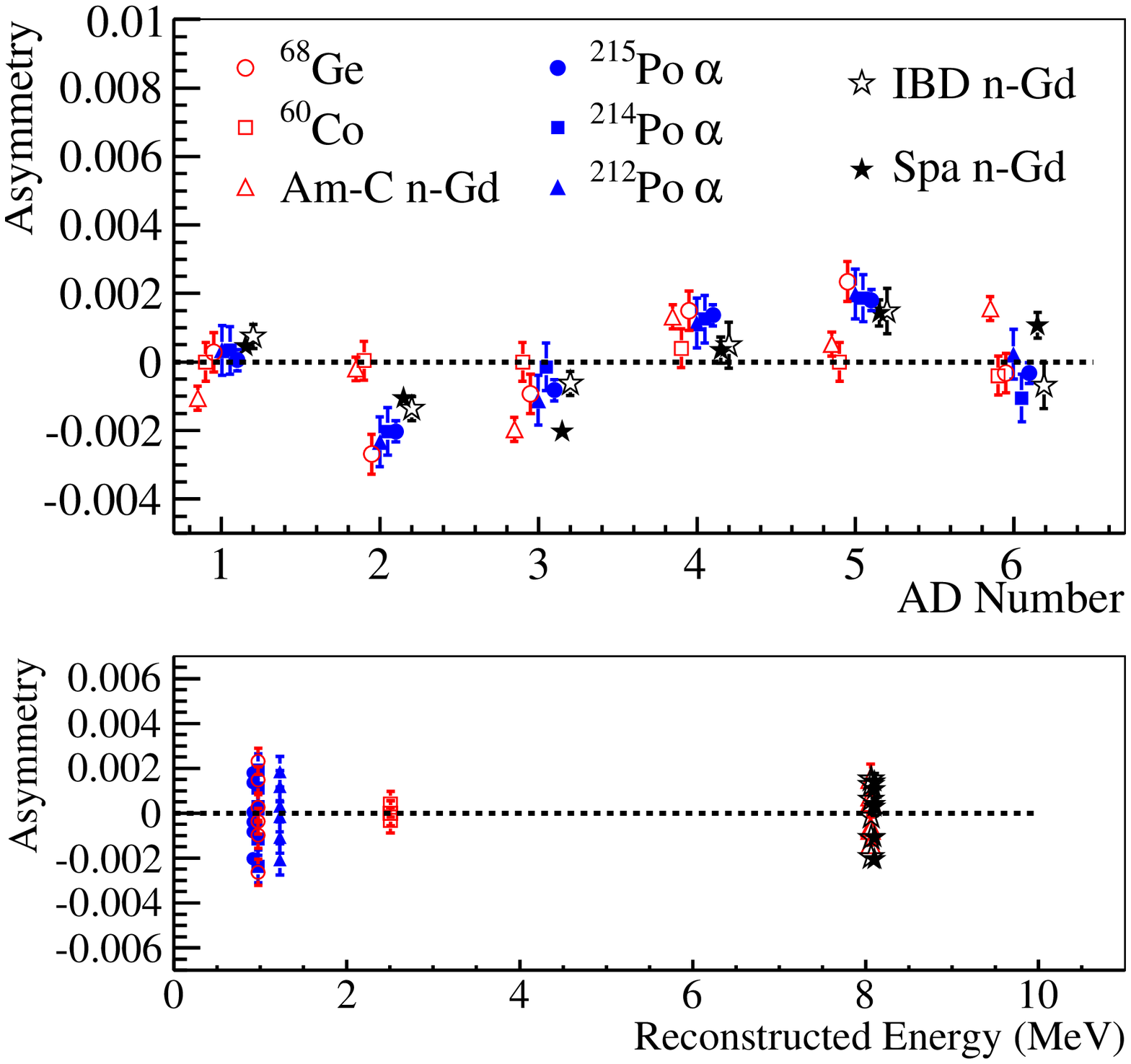}
\figcaption{Asymmetries in energy response for all six ADs. The sources $^{68}$Ge, $^{60}$Co, and Am-C were deployed at the detector center. The $^{60}$Co data was used for energy calibration. The alpha particles from Polonium decay and neutron capture on gadolinium of IBD and spallation neutrons were uniformly distributed within each detector. Differences between these sources are due to spatial non-uniformity of the detector response. The same set of data points are shown in the lower panel as a function of energy, which demonstrates that all six ADs have similar energy non-linearity.
\label{fig:s3_energy_asym}}
\end{center}
%\end{figure}

\subsection{Detector Simulation}
A Geant4~\cite{Geant4} based computer simulation (Monte Carlo, MC) of the detectors and readout electronics was used to study the detector response. It consisted of five components: kinematic generator, detector simulation, electronics simulation, trigger simulation and readout simulation.  The MC was carefully tuned to match observed detector distributions, such as PMT timing, charge response, and energy non-linearity.

\par
The antineutrino generator read from a database that stored the reactor antineutrino spectra from each core at each detector.  The database was binned in daily increments and accounted for fuel evolution. The flux was scaled later based on the actual reactor power.  The cosmic muons in the underground laboratory were simulated using a digitized topographic map of the site and Muon Simulation Code~\cite{Kudryavtsev:2008qh} (MUSIC), which calculated the energy loss and multiple scattering due to the rock overburden. The muon generator for Geant4 read randomly from a library of muon events generated with MUSIC. The software generators for the calibration sources and the simulation of the decay sequences for natural radionuclides found in our detectors were customized based on data from the ENDF database~\cite{ENDF}.

\par
All physical processes in Geant4 relevant to the Daya Bay simulation were validated. In the validation process, we found that the gamma spectra of neutron capture and muon capture on many nuclei were incorrectly modeled. Since a systematic correction was complex, we implemented corrections on a case by case basis.  The most important one was the neutron capture on gadolinium where we used a customized module based on the measured gamma spectrum~\cite{groshev}.  Furthermore, the simulation of thermal neutron scattering was improved by considering the molecular binding energy of the scattering nuclei.

\par
The gadolinium and other elemental concentrations of the liquid scintillator were measured and incorporated into the MC. All relevant optical properties of the detector components were derived from measurements, including the attenuation lengths and refractive indices of all liquids as well as the acrylic components, time constants and photon emission spectra of Gd-LS, LS, and mineral oil, and the reflectivity of the reflectors as well as other detector materials. Photon absorption and re-emission processes in liquid scintillator were modeled based on measurements in order to properly simulate the propagation of optical photons and contributions from Cherenkov process.

The details of the electronics simulation can be found in Ref.~\cite{soeren}. Using the timing and number of p.e.\ generated in PMTs, an analog signal pulse for each PMT was generated and tracked through the digitization process, taking into account the non-linearity, dark rate, pre-pulsing, after-pulsing and ringing of the waveform.  The simulated analog pulse was then used as input to a trigger system simulation for each sub-detector.

\section{Event Selection}

\subsection{IBD Selection}

Two conditions were implemented prior to the IBD selection. First, flasher events were rejected (Sec.~3.3).  Second, all AD triggers within a (-2 $\mu s$, 200 $\mu s$) time-window with respect to a water shield muon candidate ($\mu_{WS}$) were rejected, where a $\mu_{WS}$ was defined as any signal with NHIT $>$ 12 in either the inner or outer water shield. This allowed for the removal of most of the superfluous triggers that followed a muon, as well as triggers associated with muon-induced spallation products. The veto time-window was extended to 2 $\mu$s earlier than the muon to avoid time alignment issue among different detectors.  Events in an AD within $\pm$2 $\mu$s of a $\mu_{WS}$ with energy $>$20 MeV or $>$2.5 GeV were classified as AD muons ($\mu_{AD}$) or showering muons ($\mu_{sh}$), respectively.  Longer veto windows were applied for such events to further reject cosmogenic backgrounds.

The energy of the prompt and delayed candidates were required to satisfy $0.7\ \mbox{MeV}<E_{p}<12.0$~MeV and $6.0\ \mbox{MeV}<E_{d}<12.0\ \mbox{MeV}$, respectively, and $\Delta t =t_d - t_p$ must have satisfied a $1 < \Delta t < 200\,\,  \mu s$ coincidence, where $t_p$ and $t_d$ are the times of the prompt and delayed signals.  A multiplicity cut required no additional candidate with $E>0.7$ MeV in the interval 200~$\mu s$ before $t_p$, 200~$\mu s$ after $t_d$, or between $t_p$ and $t_d$.  The prompt-delayed pair was vetoed if the delayed candidate satisfied any of the conditions $-2\ \mu s< t_d - t_{\mu_{\rm WS}} \!<\! 600\ \mu$s (water shield muon), $0<t_d-t_{\mu_{\rm AD}} \!<\! 1000\ \mu$s (AD muon), or $0 < t_d - t_{\mu_{\rm sh}} \!<\! 1$~s (AD showering muon). The prompt energy, delayed energy and capture-time distributions for data and MC are shown in Figs.~\ref{fig:s4_promptcut} -~\ref{fig:s4_capturetime}, respectively.

The data are generally in good agreement with the MC. The apparent difference between data and MC in the prompt energy spectrum in Fig.~\ref{fig:s4_promptcut} is primarily due to nonlinearity of the detector response. Since all ADs had similar nonlinearity (as shown in the bottom panel of Fig.~\ref{fig:s3_energy_asym}), and the energy selection cuts cover a larger range than the actual distribution, the discrepancies between MC and data introduced negligible uncertainties to the rate analysis.  Therefore, this nonlinearity correction was not implemented in this analysis.

%\begin{figure}[htb]
\begin{center}
\includegraphics[width=\columnwidth]{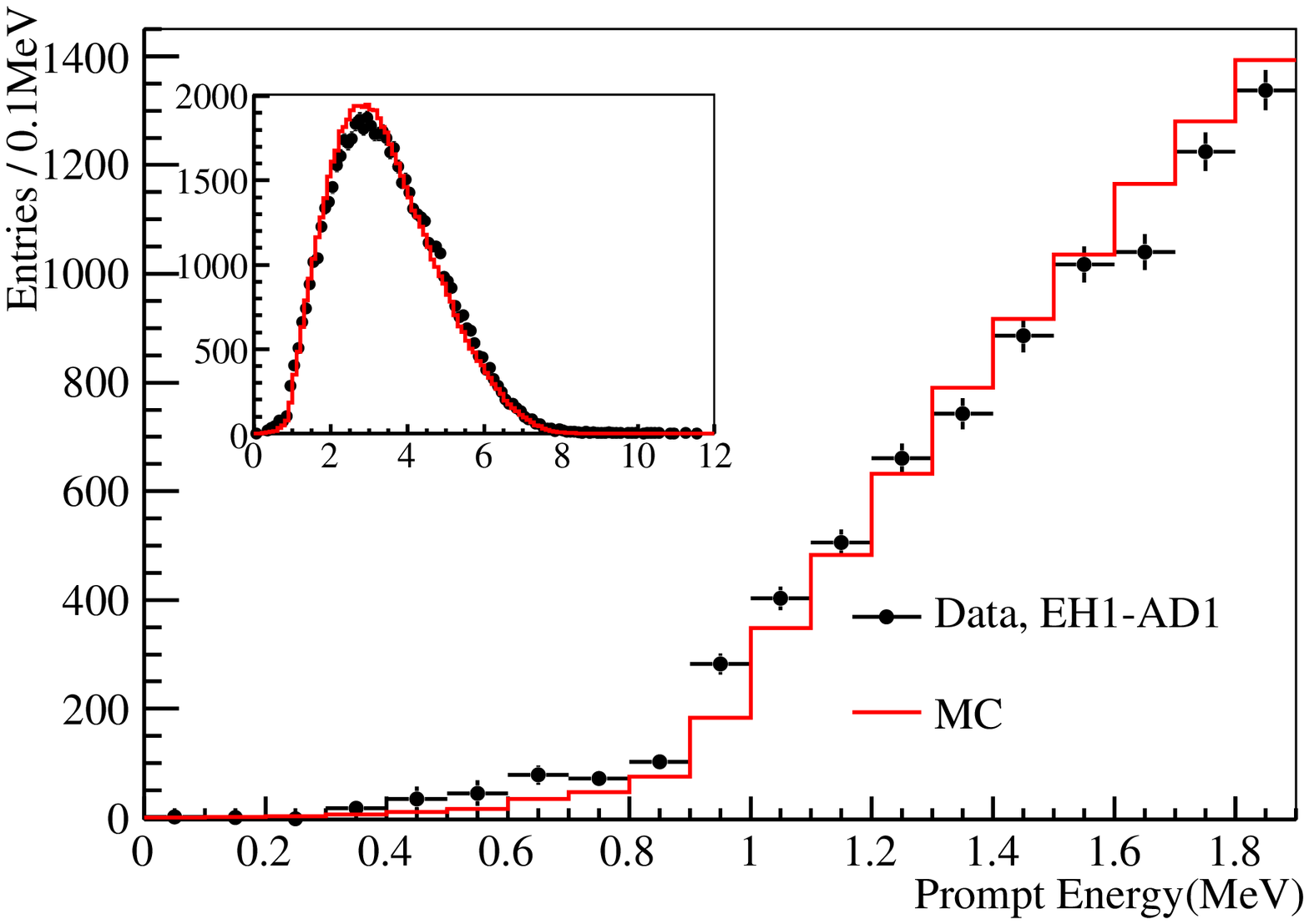}
\figcaption{Prompt energy spectrum from AD1. IBD selection required $0.7 \!<\! E_p \!<\! 12.0$ MeV. The spectrum of accidental backgrounds, determined from the distribution of all prompt-type signals, was subtracted.
\label{fig:s4_promptcut}}
\end{center}
%\end{figure}

%\begin{figure}[htb]
\begin{center}
\includegraphics[width=\columnwidth]{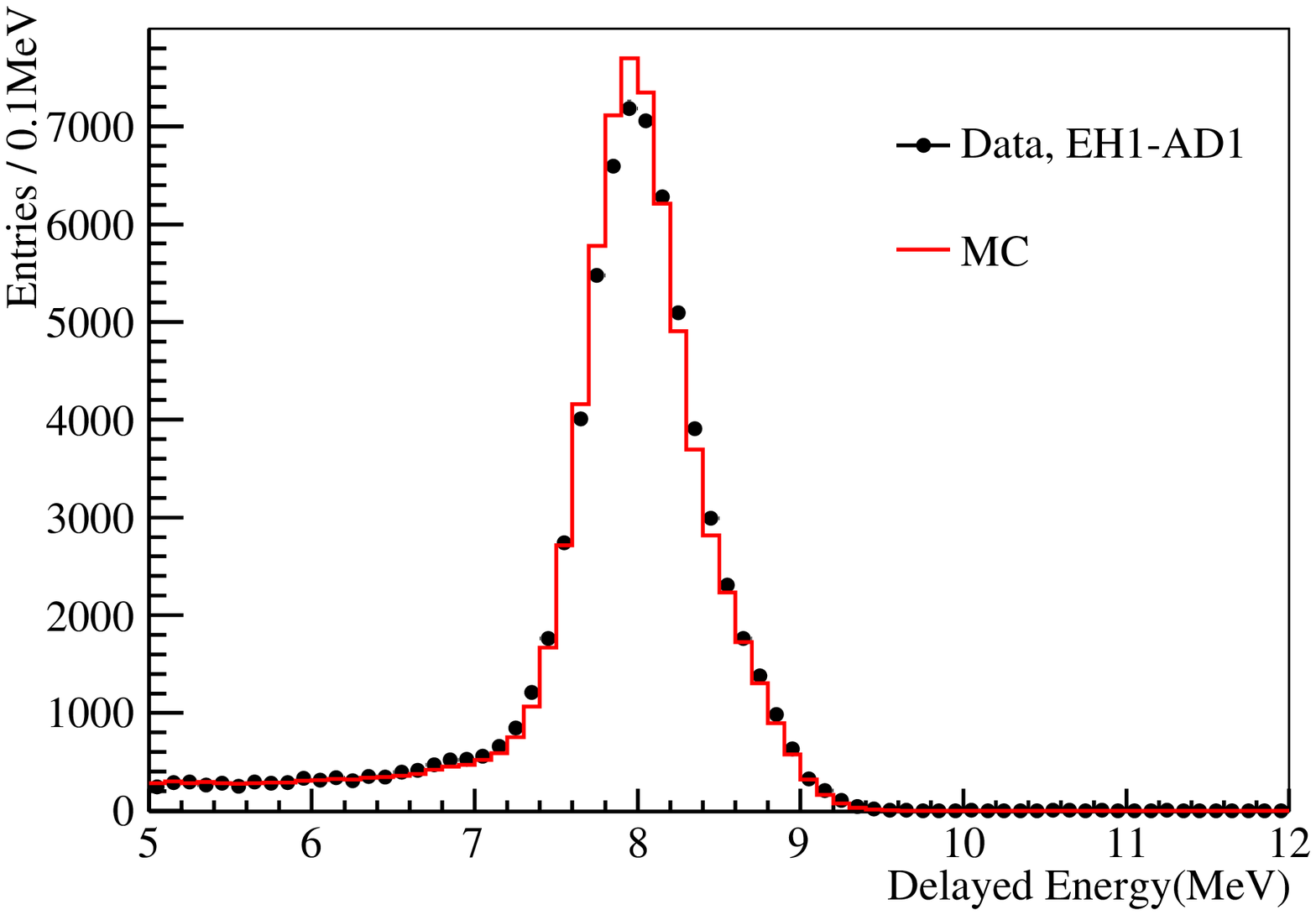}
\figcaption{Delayed energy spectrum  from AD1. IBD selection required $6.0 \!<\! E_d \!<\! 12.0$ MeV.
The spectrum of accidental backgrounds, determined from the distribution of all delayed-type signals, was subtracted.
\label{fig:s4_delaycut}}
\end{center}
%\end{figure}

\subsection{Efficiencies and Uncertainties}

For a relative measurement, the absolute efficiencies and correlated uncertainties do not factor into the error budget.  In that regard, only the relative efficiencies and uncorrelated uncertainties matter. Extraction of absolute efficiencies and correlated uncertainties was done in part to better understand our detector, and was a natural consequence of evaluating the uncorrelated uncertainties. Absolute efficiencies associated with the prompt energy, delayed energy, capture time, Gd-capture fraction, and spill-in effects were evaluated with the Monte Carlo. Efficiencies associated with the muon veto, multiplicity cut, and livetime were evaluated using data. In general, the uncorrelated uncertainties were not dependent on the details of our simulation.

%\begin{figure}[htb]
\begin{center}
\includegraphics[width=\columnwidth]{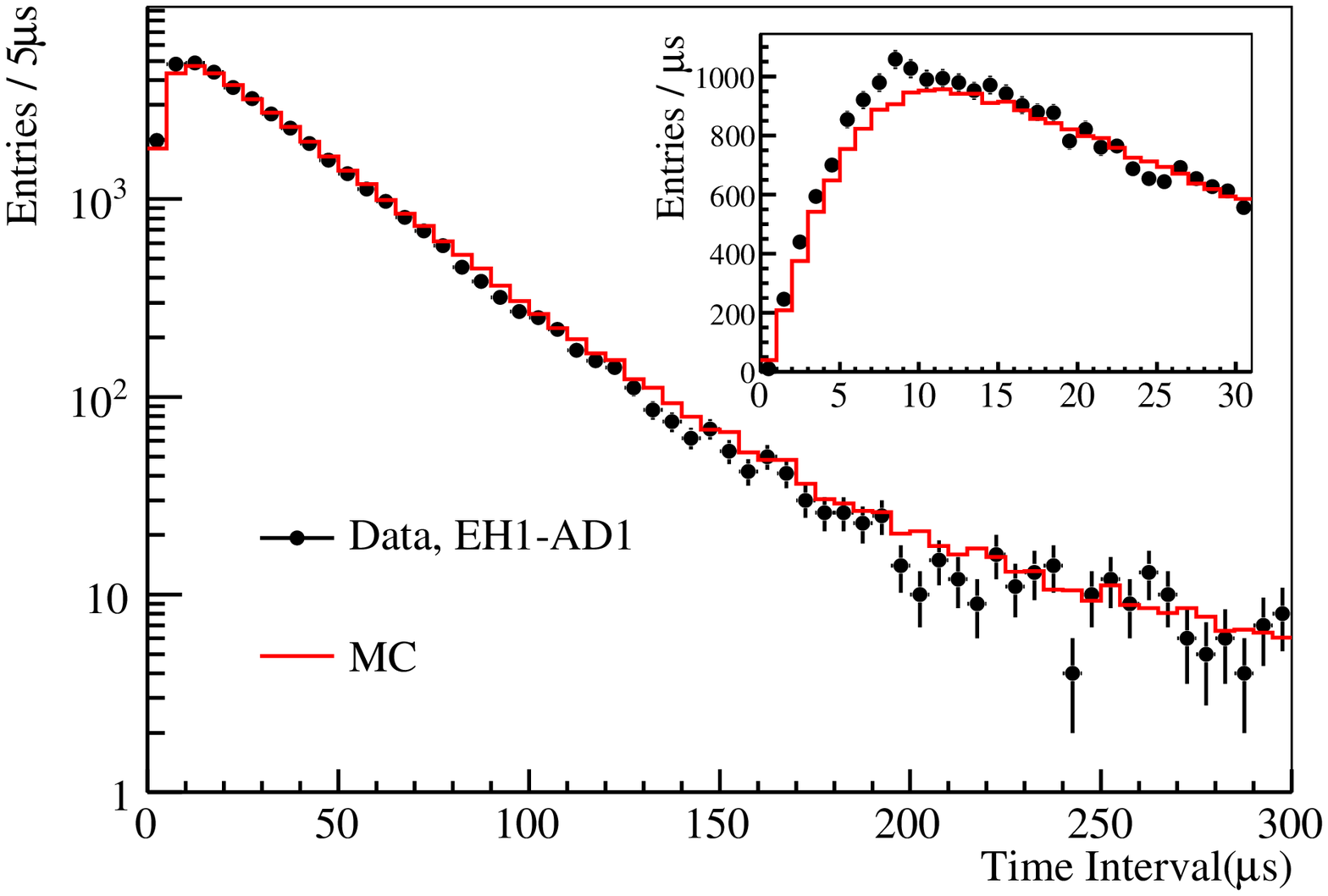}
\figcaption{Neutron capture time from AD1. IBD selection required $1 \!<\! t_d - t_p \!<\! 200$ $\mu$s.
In order to compare data with MC, a cut on the prompt energy ($E_p>3$ MeV) was applied to suppress accidental backgrounds. A zoomed-in plot for $1 \!<\! t_d - t_p \!<\! 30$ $\mu$s is shown in the inset.
\label{fig:s4_capturetime}}
\end{center}
%\end{figure}

\par
Table~\ref{tab:eff} summarizes the absolute efficiencies and the systematic uncertainties. The uncertainties of the absolute efficiencies were correlated among the ADs. No relative efficiency, except the muon veto efficiency $\epsilon_\mu$ and the average multiplicity cut efficiency $\overline{\epsilon}_m$, were corrected. All differences between the functionally identical ADs were taken as uncorrelated uncertainties.

%\begin{table}[!htb]
\begin{center}
\tabcaption{Summary of absolute efficiencies, and correlated and uncorrelated systematic uncertainties.  For our relative measurement, the absolute efficiencies as well as the correlated uncertainties effectively cancel.  Only the uncorrelated uncertainties contribute to the final error in our relative measurement.
\label{tab:eff}}
\footnotesize
\begin{tabular}{lrrr}
\toprule
  & \hspace{-0.5cm} Efficiency & \hspace{-0.2cm} Correlated & \hspace{-0.2cm} Uncorrelated \\
\hline
Target protons  &     & 0.47\% & 0.03\%  \\
Flasher cut & 99.98\% & 0.01\% & 0.01\% \\
Delayed energy cut & 90.9\% & 0.6\% & 0.12\%    \\
Prompt energy cut  & 99.88\% & 0.10\%  & 0.01\%  \\
Multiplicity cut & & 0.02\% & $<$0.01\% \\
Capture time cut  & 98.6\% & 0.12\%  &  0.01\%  \\
Gd capture fraction  & 83.8\% & 0.8\% & $<$0.1\%  \\
Spill-in     & 105.0\% & 1.5\% & 0.02\%    \\
Livetime & 100.0\% & 0.002\% & $<$0.01\% \\
\hline
Combined & 78.8\%  & 1.9\%   & 0.2\% \\
\bottomrule
\end{tabular}
\end{center}
%\end{table}

\par
The absolute efficiency of the prompt energy cut ($0.7<E_p<12.0$ MeV) was determined to be 99.88\%.  The energy spectrum is shown in Fig.~\ref{fig:s4_promptcut}. Inefficiency was mainly caused by interactions inside the inner acrylic vessel, indicated by the vertex distribution of the rejected prompt signal below 0.7 MeV. While the uncertainty in the energy scale is below 0.5\% for events at the detector center or uniformly distributed in the target volume, it is larger for events at the edge. The asymmetries of the energy among ADs could be as large as 2\% for events at the radius of ACU-C, studied with $^{60}$Co source~\cite{ad12}.
Taking 2\% uncertainty in the energy scale for events near the inner acrylic vessel, the uncorrelated uncertainty of the efficiency due to the prompt energy cut was evaluated to be 0.01\%. Given that the positron threshold was calibrated with the $^{68}$Ge source, the uncertainty of this absolute efficiency comes from the difference of non-linearity and non-uniformity between the data and MC. The correlated uncertainty was estimated to be 0.1\%.

\par
The absolute efficiency of the delayed energy cut ($6.0<E_d<12.0$ MeV) was determined to be 90.9\%.
As shown in Fig.~\ref{fig:s4_delaycut}, the fraction of events in the 6-7 MeV region was 5.3\% of that in 6-12 MeV for MC.  For selected IBD data, this fraction was 5.6\%. Assuming the same relative difference between MC and data in the 0-6 MeV region, the difference of absolute efficiency between the MC and data was evaluated to be 0.6\%, which is taken as the correlated uncertainty.
By varying the cut at 6 MeV and counting the number of events in the selected sample, we found that the 0.5\% asymmetry of the energy scale in ADs, shown in Fig.~\ref{fig:s3_energy_asym}, leads to a 0.12\% uncorrelated efficiency uncertainty. The low energy tail around 6 MeV is relatively flat and the MC and data agree well. Both MC and data studies yield the same uncorrelated efficiency uncertainty.

\par
The spill-in enhancement resulted when neutrons from IBD interactions outside the target volume were captured by a Gd nucleus in the target volume. It was defined as the ratio of all IBD interactions that lead to a neutron capture on Gd to IBD interactions within the target volume leading to a neutron capture on Gd.  From MC, it was evaluated to be 105.0\%. By modeling the relative difference in acrylic vessel thickness, acrylic density and liquid density in MC, the relative uncertainty of the spill-in efficiency was evaluated to be 0.02\%. The correlated uncertainty of the spill-in efficiency was evaluated with MC. The modeling of molecular binding energy of the scattering nuclei has a large impact on the simulation of thermal neutron scattering, and thus on the absolute spill-in efficiency. The thermal neutron scattering process is correlated with the neutron capture time. The agreement between data and MC is shown in the inset of Fig.~\ref{fig:s4_capturetime}. By comparing the results of simulation with two different models of molecular binding energy as well as without binding energy, we conservatively estimated the correlated uncertainty of the spill-in efficiency to be 1.5\%.

\par
The Gd capture fraction was defined as the ratio of the number of Gd capture events produced by IBD reactions to all IBD reactions in the Gd-LS.  It was evaluated to be 83.8\%. The spill-out deficit, $\sim$2.2\% by comparing the Gd capture fraction of the Am-C neutron source at the detector center and IBD events in MC, was included in the absolute Gd capture fraction. Spill-out is analogous to spill-in, except that IBD neutrons produced within the target volume were captured outside the target volume. By measuring the difference in the neutron capture time of each AD, the relative Gd-concentration variation was constrained and the Gd capture fraction variation was determined to be within 0.1\%. By comparing Am-C source data with MC, as well as spallation neutrons, the correlated uncertainty on Gd capture fraction was estimated to be 0.8\%.

\par
The efficiency of the capture time cut ($1<\Delta t <200$ $\mu$s) was evaluated to be 98.6\% with 0.2\% of events with $\Delta t<1\ \mu s$ and 1.2\% events with $\Delta t>200\ \mu s$. The correlated efficiency uncertainty was evaluated to be 0.12\%, according to the difference in the measured capture time between Am-C data and MC. The uncorrelated uncertainty comes from the Gd-concentration variation and possible trigger time-walk effect, and it was evaluated to be 0.01\%.

\par
The muon veto efficiencies were determined using data. For each type of muon candidate ($\mu_{WS}$, $\mu_{AD}$ and $\mu_{sh}$), the start and end time of the veto window were well defined. Overlapping veto windows were merged to avoid double counting.  As a result each livetime window was precisely calculated as the unvetoed time interval between two isolated veto windows. The total livetime was obtained by summing all the individual livetime windows.  The muon veto efficiency $\epsilon_\mu$ was defined as the fraction of the livetime after a muon veto in the total DAQ livetime. For each experimental hall the muon rates were stable as shown in Fig.~\ref{fig:s3_WSTrigger}. The muon veto efficiencies differed due to different muon candidate rates.

\par
The multiplicity cut required no additional $>$~0.7~MeV signals (singles) in the time range from 200~$\mu$s before the prompt signal to 200~$\mu$s after the delayed signal. The singles rate $R_s$ can be taken as the rate of prompt-type signals shown in Fig.~\ref{fig:s3_singles}.
The efficiency of the multiplicity cut is a product of three components. The probability of no singles in the 200~$\mu$s before the prompt signal is given by $\exp(-R_{200})$, where $R_{200}=R_s\cdot200\mu s$.  The probability of no singles between the prompt and delayed signal is given as $\int_0^{200\mu s} \exp({-R_s t}) f(t) dt$, where $f(t)$ is the probability density function of the capture time of neutron on Gd, and can be simplified as $1-R_s\overline{t}_{\rm cap}+O(10^{-5})$, where $\overline{t}_{\rm cap}$ is the mean neutron capture time in 200~$\mu$s. The average of the mean capture time of the six ADs was 33.46 $\mu$s, obtained from data. The uncorrelated uncertainty was determined by the difference of the mean capture times among ADs. The probability of no singles in 200~$\mu$s after the delayed signal must be calculated for two cases since the window may be truncated by an AD muon that would obscure any potential single. If the single livetime window was $T_s<200\mu$s, the efficiency was
\begin{equation}
\dfrac{1-e^{(-R_s T_s)}}{R_s T_s}\,,\nonumber
\end{equation}
and if $T_s>200\mu$s, the efficiency was
\begin{equation}
(1-\dfrac{200\mu s}{ T_s})e^{-R_{200}}  + \dfrac{1}{R_s T_s}(1-e^{-R_{200}} )\,. \nonumber
\end{equation}
Because the second term depends on the length of the single livetime window, the multiplicity cut efficiency must be calculated for every single livetime window. As a consequence, the muon veto efficiency and the multiplicity were coupled. The combined efficiency is
\begin{equation}
  \epsilon_\mu \overline{\epsilon}_m = \left(\sum_i {\epsilon_m^i T_s^i} \right)/T_{\rm DAQ}\,,
\label{eq:multi}
\end{equation}
where $\epsilon_m^i$ is the multiplicity cut efficiency in the $i$-th single livetime $T_s^i$, and $T_{\rm DAQ}$ is the analyzed good run time.
The muon veto efficiency $\epsilon_\mu$ and the average multiplicity cut efficiency $\overline{\epsilon}_m$ calculated with Eq.~\ref{eq:multi} are listed in Table~\ref{tab:ibd} and corrected for each AD.

\par
The target mass uncertainty was discussed extensively in Ref.~\cite{ad12}. The correlated uncertainty 0.47\% largely comes from the hydrogen-carbon ratio of the target liquid, which is canceled out in the near-far relative measurement by using the same batch of Gd-LS. The time variation of the target mass, e.g.\ due to temperature variation, is monitored by the liquid level with several independent sensors in the overflow tanks on the top of the AD lid~\cite{ADlid}. The variation of the target mass for the analyzed data set is shown in Fig.~\ref{fig:s4_mass}. The $\pm 0.02$\% range is the target mass uncertainty evaluated during filling~\cite{ad12}. To accurately evaluate the mass of Gd-LS transferred into detectors, a 20-t filling tank was equipped with load cells to measure the mass of the filling tank before and after filling. The above uncertainty is dominated by the load cell drift during the filling operation. As such, the uncorrelated uncertainty is set to be 0.03\%.

%\begin{figure}[htb]
\begin{center}
\includegraphics[width=\columnwidth]{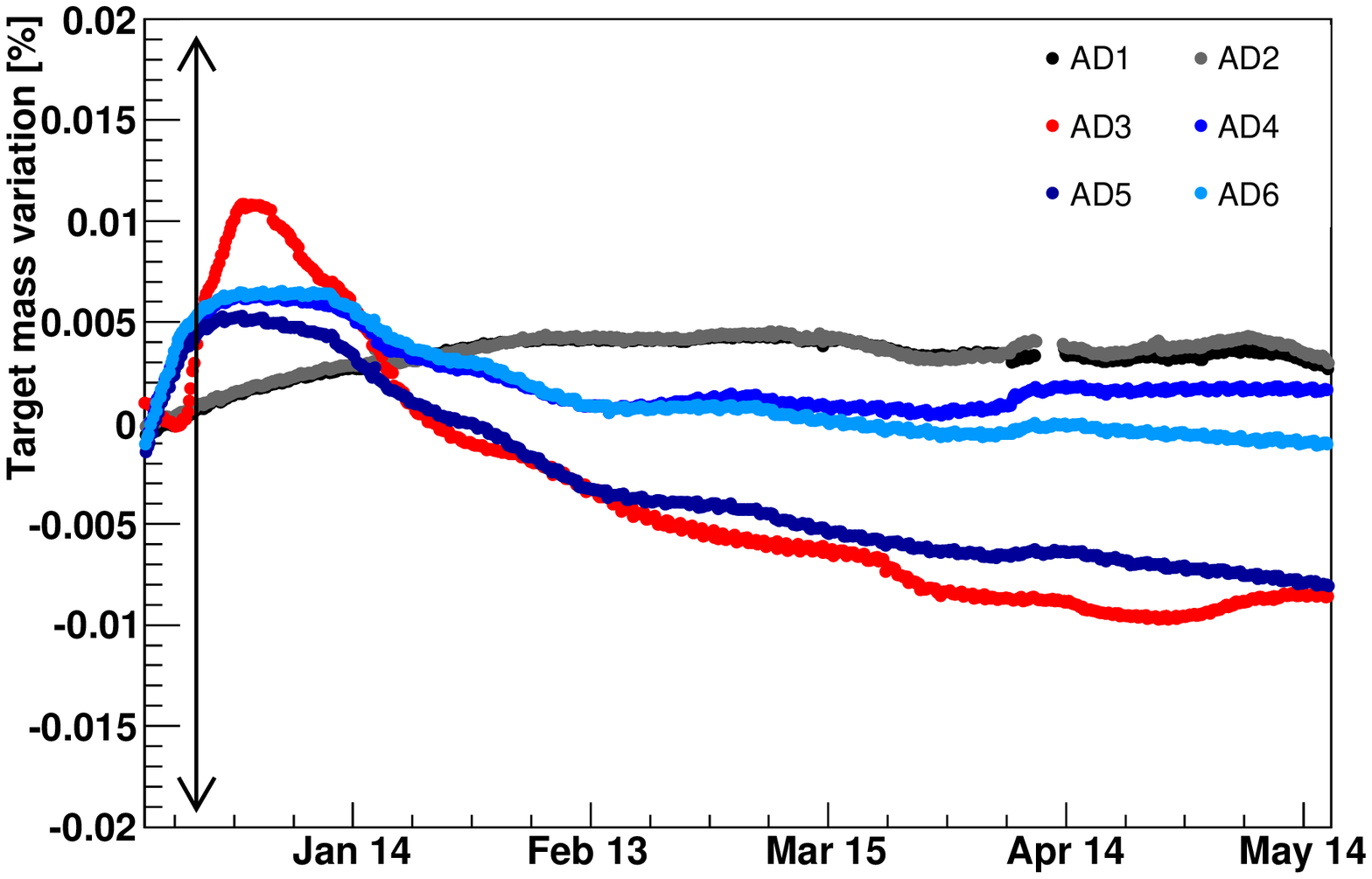}
\figcaption{Target mass variation for each AD over the analyzed time period. The vertical double arrow indicates the total uncorrelated uncertainty in the target mass evaluated during filling.
\label{fig:s4_mass}}
\end{center}
%\end{figure}

\section{Backgrounds}
\subsection{Accidental Backgrounds}

The accidental background is defined as any pair of otherwise uncorrelated signals that happen to satisfy the IBD selection criteria. For any given signal with an observed energy between 6 and 12 MeV (delayed-type signal), the probability of forming an accidental background is the product of two components, the probability of a prompt-type signal within 1-200 $\mu s$ before the delayed-type signal, $1-\exp(-R\cdot199\mu s)$, and the probability of no singles within 200$\mu s$ before the prompt-type signal and 200 $\mu s$ after the delayed-type signal, $\exp(-R\cdot400\mu s)$.   $R$ is the rate of prompt-type singles.  Since the rate of prompt-type and delayed-type singles changed over time, the accidental background was calculated every four hours and summed as follows:
\begin{equation}
\label{eqn_acc}
  N_{\rm acc.bkg.}=\sum_i N_i e^{-R_i\cdot 400\mu s} (1-e^{-R_i\cdot199\mu s}) \,,
\end{equation}
where $N_i$ and $R_i$ are the number of delayed-type and prompt-type singles rates in the $i$-th four-hour period, respectively. The statistical uncertainty was dominated by $N_i$, and was approximated as
\begin{equation}
  \delta N_{\rm acc.bkg.}^{\rm (stat.)} \approx \dfrac{N_{\rm acc.bkg.}}{\sqrt{\sum_i N_i}} \,.
\end{equation}
The expected rates of accidental backgrounds are listed in Table~\ref{tab:ibd}, after correcting for the muon veto efficiency and the multiplicity cut efficiency in the IBD selection.

\par
An alternate method to determine the accidental backgrounds, the off-window method, was developed. By definition, the accidental background within the IBD coincidence time window  ($1\ {\mu}s<{\Delta}t<200\ {\mu}s$) should be the same as in any other window ($t^{\rm off}+1\ {\mu}s<{\Delta}t<t^{\rm off}+200\ {\mu}s$), where $t^{\rm off}$ is an arbitrary time offset. If $t^{\rm off}$ is large enough to avoid real correlated events (such as for IBD, fast neutron (Sec.~5.2), and $^{9}\mbox{Li}/^{8}\mbox{He}$ decay (Sec.~5.3)), the accidental backgrounds can be estimated by counting the coincidences in the off-window. To reduce the statistical uncertainty, multiple non-overlapping off-windows were examined. The mean number of selected coincidences in these off-windows was taken as the expected accidental background. The relative differences between the results from the off-window method and the calculations using Eq.~\ref{eqn_acc} were consistent given the statistical uncertainties for all six ADs.

\par
The accidental background was also validated by comparing the distributions of distance between the reconstructed vertices for the prompt and delayed signals of the IBD candidates and accidentals selected by the off-window method, as shown in Fig.~\ref{fig:s5_accidental}. The prompt and delayed vertices of accidentals were uncorrelated, thus giving a broad distribution, while the two vertices for IBD events were correlated, giving a distribution peaked at short distance. For distances greater than 2 m, the IBD candidate and off-window distributions agree well.

%\end{multicols}
%\vspace{1.0cm}
\begin{table*}
\begin{center}
\caption{Summary of signal and background. The background and IBD rates have been corrected for the muon veto efficiency $\epsilon_\mu$ and the average multiplicity cut efficiency $\overline{\epsilon}_{m}$. \label{tab:ibd}  }
\footnotesize
\begin{tabular}{ccccccc}
\toprule
                  & AD1  & AD2  & AD3 & AD4 & AD5 & AD6 \\
\hline
IBD candidates &  69121  &  69714  &  66473 &  9788  &  9669  &  9452 \\
%\hline
Expected IBDs & 68613 & 69595 & 66402  & 9922.9 &  9940.2 &  9837.7 \\
DAQ livetime (days) & \multicolumn{2}{c}{127.5470}     &  127.3763  &    &   126.2646  &    \\
%\hline
$\epsilon_\mu$ &  0.8231  &  0.8198  &  0.8576  &  0.9813  &  0.9813  & 0.9810  \\
%\hline
$\overline{\epsilon}_{m}$ &  0.9738 &  0.9742  &  0.9753  & 0.9737  &  0.9734  &  0.9732 \\
%\hline
Accidentals (per day) &  9.73$\pm$0.10  &  9.61$\pm$0.10  & 7.55$\pm$0.08   &  3.05 $\pm$0.04  &  3.04 $\pm$ 0.04  &  2.93 $\pm$0.03 \\
%\hline
Fast-neutron (per day) &  0.77$\pm$0.24  &  0.77$\pm$0.24   &  0.58$\pm$0.33   &  0.05$\pm$0.02  &
 0.05$\pm$0.02 & 0.05$\pm$0.02 \\
%\hline
$^9$Li/$^8$He (per AD per day) &  \multicolumn{2}{c}{2.9$\pm$1.5}  & 2.0$\pm$1.1 & \multicolumn{3}{c}{0.22$\pm$0.12}   \\
%\hline
Am-C correlated (per AD per day) &  \multicolumn{6}{c}{0.2$\pm$0.2}   \\
%\hline
 ($\alpha$, n) background (per day)&  0.08$\pm$0.04  &  0.07$\pm$0.04  & 0.05$\pm$0.03 & 0.04$\pm$0.02  & 0.04$\pm$0.02 & 0.04$\pm$0.02   \\
%\hline
\hline
IBD rate (per day) &  662.47$\pm$3.00  & 670.87$\pm$3.01 & 613.53$\pm$2.69  & 77.57$\pm$0.85  & 76.62$\pm$0.85  & 74.97$\pm$0.84 \\
\bottomrule
\end{tabular}
\end{center}
\end{table*}
%\vspace{1.0cm}
%\begin{multicols}{2}

\subsection{Fast Neutron Backgrounds}

Energetic neutrons created by cosmic rays entering an AD could mimic IBD by recoiling off a proton before being captured on Gd.  Since the visible energy of the recoil proton ranged well past that of the IBD events (up to several hundred MeV as shown in Fig.~\ref{fig:s5_fastn}), we estimated the number of fast-neutron background events in the IBD sample by extrapolating the prompt energy ($E_p$) distribution between 12 and 100~MeV down to 0.7~MeV. Two different extrapolation methods were used. By assuming the recoil proton energy spectrum follows a flat distribution, the mean number of events per energy bin of the distribution from 12 to 100~MeV was used to estimate the number of fast-neutron events between 0.7 and 12 MeV.  Alternately, the data from 12 to 100~MeV were fit with a first-order polynomial function ($f(E)=a+b E$).  The best-fit parameters were used to estimate the number of fast-neutron events between 0.7 and 12~MeV.  The fast neutron background in the IBD sample was assigned to be equal to the mean value of the two extrapolation methods. The systematic error was determined from their differences and the fitting uncertainties.

%\begin{figure}[htb]
\begin{center}
\includegraphics[width=\columnwidth]{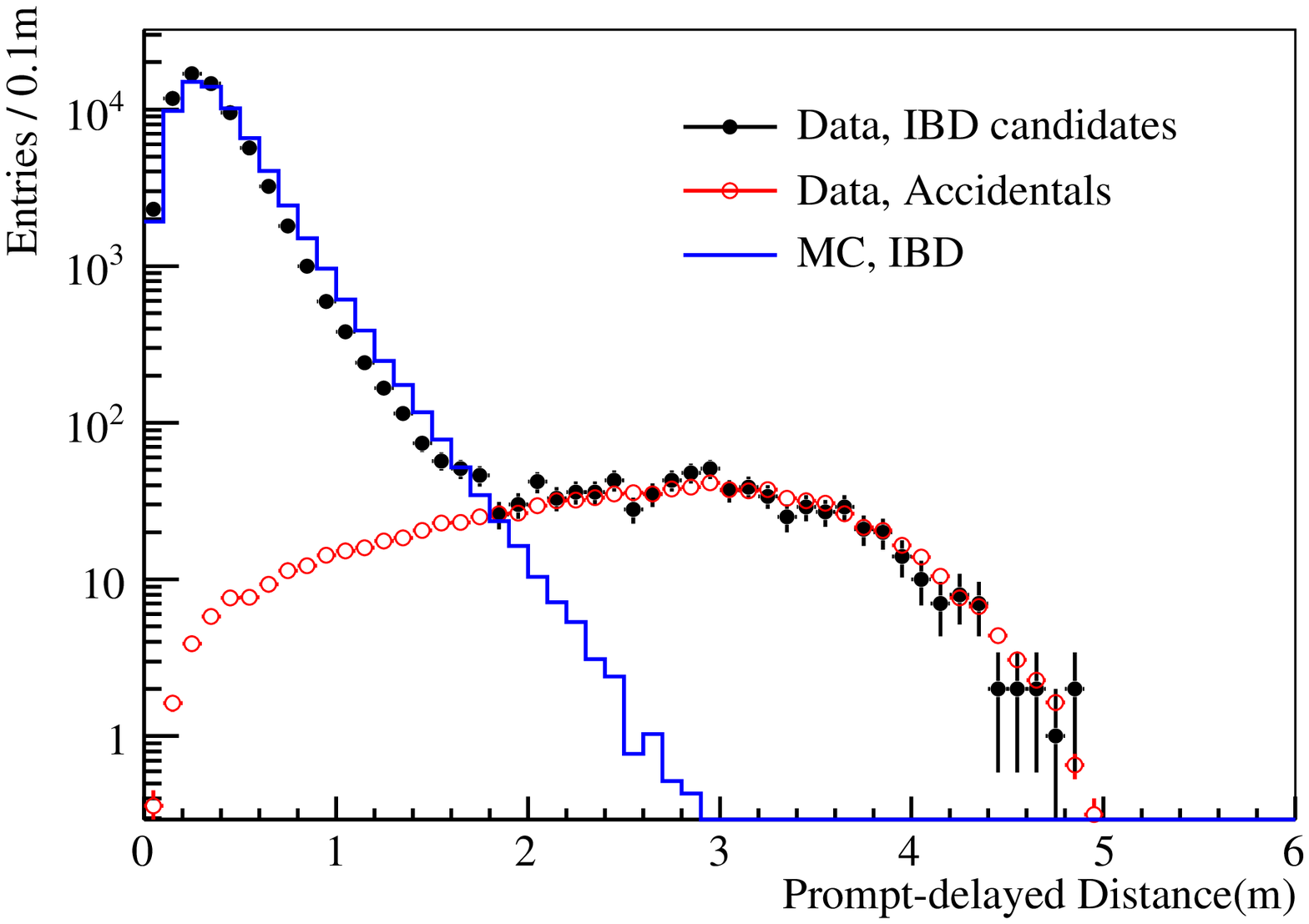}
\figcaption{Distance between the prompt signal and delayed signal. The dots show the IBD candidates in data and the open circles are accidental candidates selected with the off-window method, both in their absolute rates. The histogram shows the simulated IBD events, with rate normalized to data. \label{fig:s5_accidental}}
\end{center}
%\end{figure}

\par
As a check, we studied the fast neutrons associated with tagged muons. The prompt energy of the fast neutron tagged by the IWS muon will be contaminated if the muon clips the edge or corner of an AD. Furthermore, the fast neutron backgrounds in the IBD candidate pool mostly originated from OWS muons (defined as OWS PMT multiplicity $>12$ and without an IWS trigger) or muons passing through rock, since the muon detection efficiency of the IWS was very high (99.7\%). The fast neutrons tagged by the OWS muons or RPC-only muons (only detected by RPC) had a prompt energy spectrum similar to the fast neutron backgrounds in the IBD sample.  After rejecting flasher events, we selected fast-neutron-like events by requiring exactly two signals within 200~$\mu$s after an OWS muon or a RPC-only muon. The time interval and the energy selections of the prompt-delayed pair were the same as the IBD selections, except that the prompt energy was relaxed to be 0.7 $<\ E_p\ <$ 100 MeV. We combined the samples from EH1 and EH2 to create the fast neutron prompt energy spectrum shown in the inset of Fig.\ref{fig:s5_fastn}. The observed distribution validates our extrapolation method for estimating the fast neutron background.

%\begin{figure}[htb]
\begin{center}
\includegraphics[width=\columnwidth]{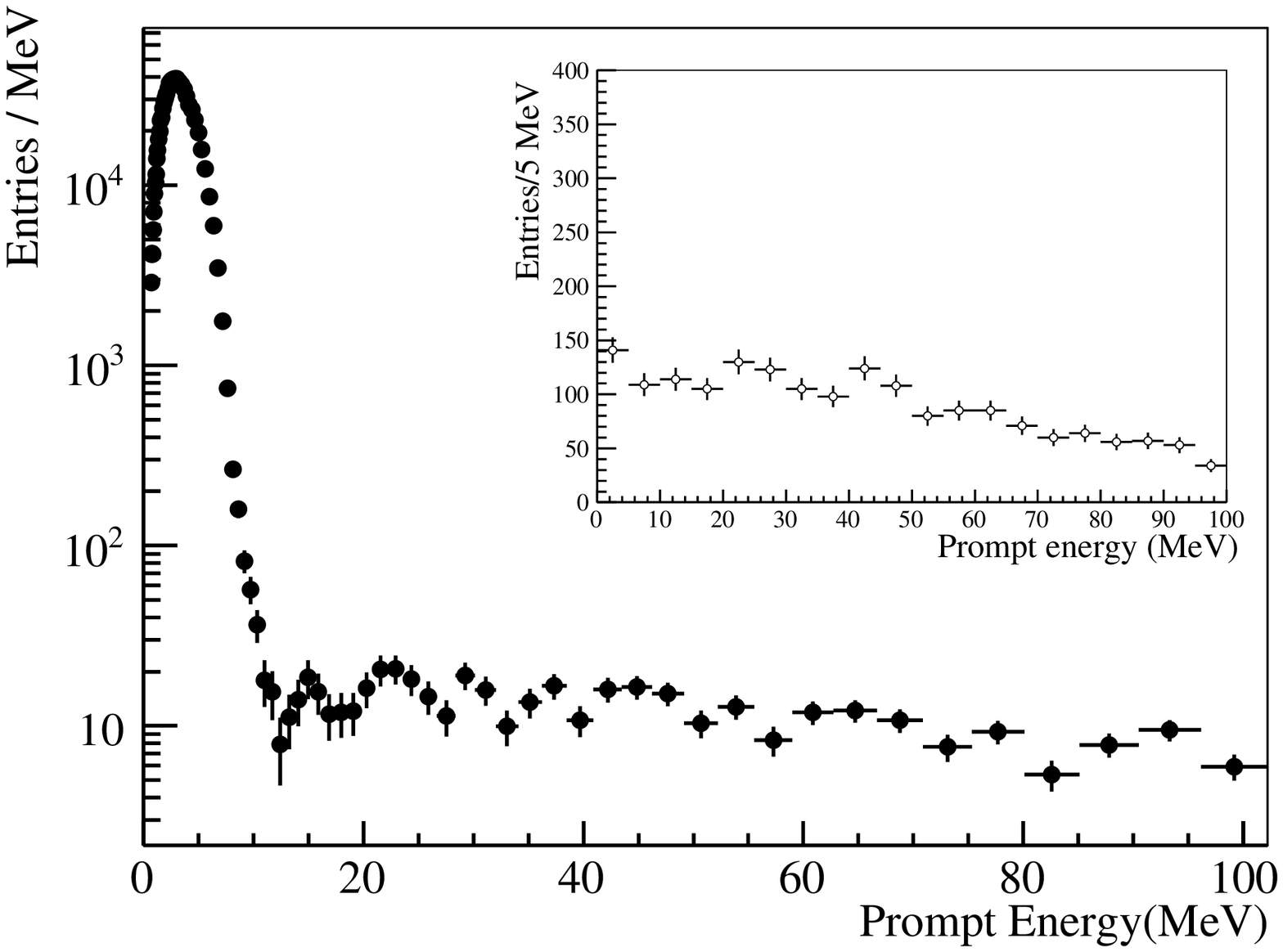}
\figcaption{Prompt energy spectrum of IBD candidates with the upper limit relaxed. The energy spectrum of the fast neutron backgrounds tagged by the OWS muons or RPC-only muons is shown in the inset.\label{fig:s5_fastn}}
\end{center}
%\end{figure}

\par
Two additional methods were used to provide further cross checks and estimates for the fast neutron background. These two methods are consistent with the result by extrapolating the IBD prompt energy spectrum within the assigned uncertainty.

\par
First, a muon with a large PMT multiplicity in the IWS and OWS has an increased probability to produce a fast neutron in an AD, presumably since track length correlates with PMT multiplicity.  Such a correlation has been observed in data.  Also, muon detection inefficiency was associated with low PMT multiplicity. By extrapolating the fast neutron rate produced by muons with a sum of PMT multiplicities between 24 and 48 in the IWS and OWS to the range 0 and 24, we were able to estimate the fast neutron background slipping into the IBD sample due to the inefficiency of the muon detection.

\par
Second, we collected different fast neutron samples based on muons going through different detector volumes ($n^{\rm IWS}_{f}$: fast neutron from an IWS tagged muon; $n^{\rm OWS}_{f}$: fast neutron from an OWS muon; $n^{\rm rock}_{f}$: fast neutron from a muon going through nearby rock) and estimated these samples separately. The $n^{\rm rock}_{f}$ was estimated by selecting RPC-only muons.  MC simulation suggested that the fast neutron backgrounds tagged by RPC-only muons account for one-third of the rock neutron background.  The fast neutron background ($n_{f}$) is described as
\begin{eqnarray}  \label{eqn:fastn}
n_{f}=n^{\rm IWS}_{f}(1-\xi_{\rm IWS})+n^{\rm OWS}_{f}(1-\xi_{\rm OWS})+n^{\rm rock}_{f}
\end{eqnarray}
where $\xi_{\rm IWS}$ is the muon detection efficiency of the IWS and $\xi_{\rm OWS}$ is that of the OWS.

\subsection{$^9$Li/$^8$He Backgrounds}

The rate of correlated background from the $\beta$-n cascade of the cosmogenic $^9$Li/$^8$He decays was evaluated from the distribution of the time since the last muon, which can be described as~\cite{wenljnim}
\begin{equation}\label{eqn:li9FitFunc}
f(t) = \frac{B_a} {\lambda_a} \cdot e^{-t/\lambda_a } +\frac{B_b}{\lambda_b} \cdot e^{-t/\lambda_b} + \frac{N_{IBD}}{T} e^{-t/T}\,,
\end{equation}
\noindent where $B_a$ and $B_b$ are the number of $\beta$-n events for $^9$Li and $^8$He, respectively. $T$ is the mean time between muons, $\frac{1}{\lambda_a}=\frac{1}{T}+\frac{1}{\tau_a}$ and $\frac{1}{\lambda_b}=\frac{1}{T}+\frac{1}{\tau_b}$ with $\tau_a=0.257$ s and $\tau_b=0.172$ s being the known decay time constants for $^9$Li and $^8$He, respectively. The muon rate $R_\mu = 1/T$ depends on the muon selection criteria.

To reduce the impact of accidental backgrounds on our measurement of $^9$Li and $^8$He, we made the following modification to our IBD selection criteria:
\begin{itemize}
  \item $0.7 <E_p< 12.0$ MeV changed to $3.5 <E_p< 12.0$ MeV.
  \item $1 < \Delta t < 200\mu s$ changed to $1 < \Delta t < 100 \mu s$.
  \item  $\mu_{sh}$ veto time changed from 1 s to 1000 $\mu$s
\end{itemize}
The measured $^9$Li/$^8$He rate was corrected for the relative efficiency with respect to the IBD selection criteria.  Assuming that $^9$Li was predominant over $^8$He (as observed in a previous experiment~\cite{KamLAND} and consistent with our observations), and based on the $^9$Li $\beta$ spectrum, this efficiency was evaluated to be about 72\%.  The reduced capture time window has an efficiency of 94\%.  The residual accidental backgrounds was thus reduced to $< 0.05$/day at the near sites, and $< 0.01$/day at the far site.

\par
To reduce the number of minimum ionizing muons in these data samples, we assumed that most of the $^9$Li and $^8$He production was accompanied with neutron generation, and thus rejected AD tagged muon events with no follow-on neutron (defined as $>$ 1.8 MeV signal within a 10 - 200 $\mu$s window). The muon samples with and without reduction were both prepared for the $^9$Li and $^8$He background estimation.  The data was sub-divided into six groups in visible muon energy (0.02-0.5, 0.5-1.5, 1.5-2.5, 2.5-3.5, 3.5-4.5, and $>$ 4.5 GeV). Taking EH1 as an example, the corresponding muon rates in each energy bin were (10.0, 10.9, 0.23, 0.042, 0.016, 5.6e-3 Hz). Note that the maximum visible energy was around 5 GeV because of the saturation of the PMTs. An example of a fit to the time-since-last muon distribution using Eq.~\ref{eqn:li9FitFunc} for determining the number of $^9$Li and $^8$He events for $E_\mu>$ 4.5 GeV is shown in Fig.~\ref{fig:s5_lifit}. Though only four seconds are shown in the figure, the fit range was actually from 1 ms to 40 s. Fitting over such a large range helped to insure that $R_\mu$ was accurate. Because of the 1000 $\mu$s $\mu_{AD}$ veto, the fitted $R_\mu$ was slightly smaller than the directly measured value.

%\begin{figure}[htb]
\begin{center}
\includegraphics[width=\columnwidth]{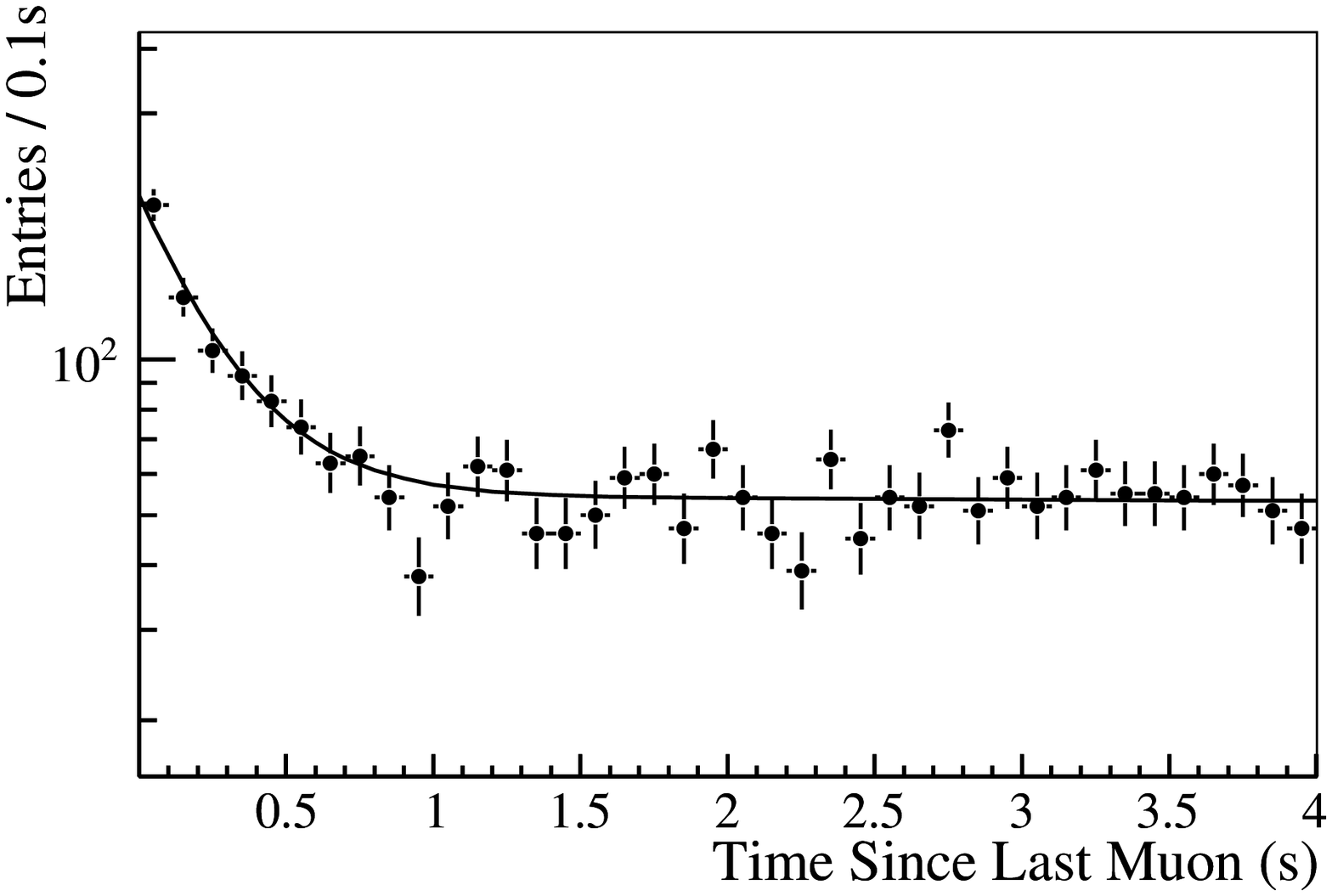}
\figcaption{An example of fitting for $^8$He/$^9$Li backgrounds.
\label{fig:s5_lifit}}
\end{center}
%\end{figure}

\par
Instead of allowing the $^9$Li to $^9$Li plus $^8$He ratio to float, we scanned it from 0 to 1 in steps of 0.01. For each scan point, a maximum likelihood fit was done, where only $N_{Li+He}$, $N_{IBD}$, and $R_\mu$ were allowed to float. Also, only the results with a global maximum likelihood among scan points were regarded as best fit values. The global maximum likelihood confirmed that $^9$Li was dominant in the $^8$He/$^9$Li backgrounds.  The binning effect was included in the uncertainty estimation by changing the bin width of the time-since-last muon distribution.

%\begin{figure}[htb]
\begin{center}
\includegraphics[width=\columnwidth]{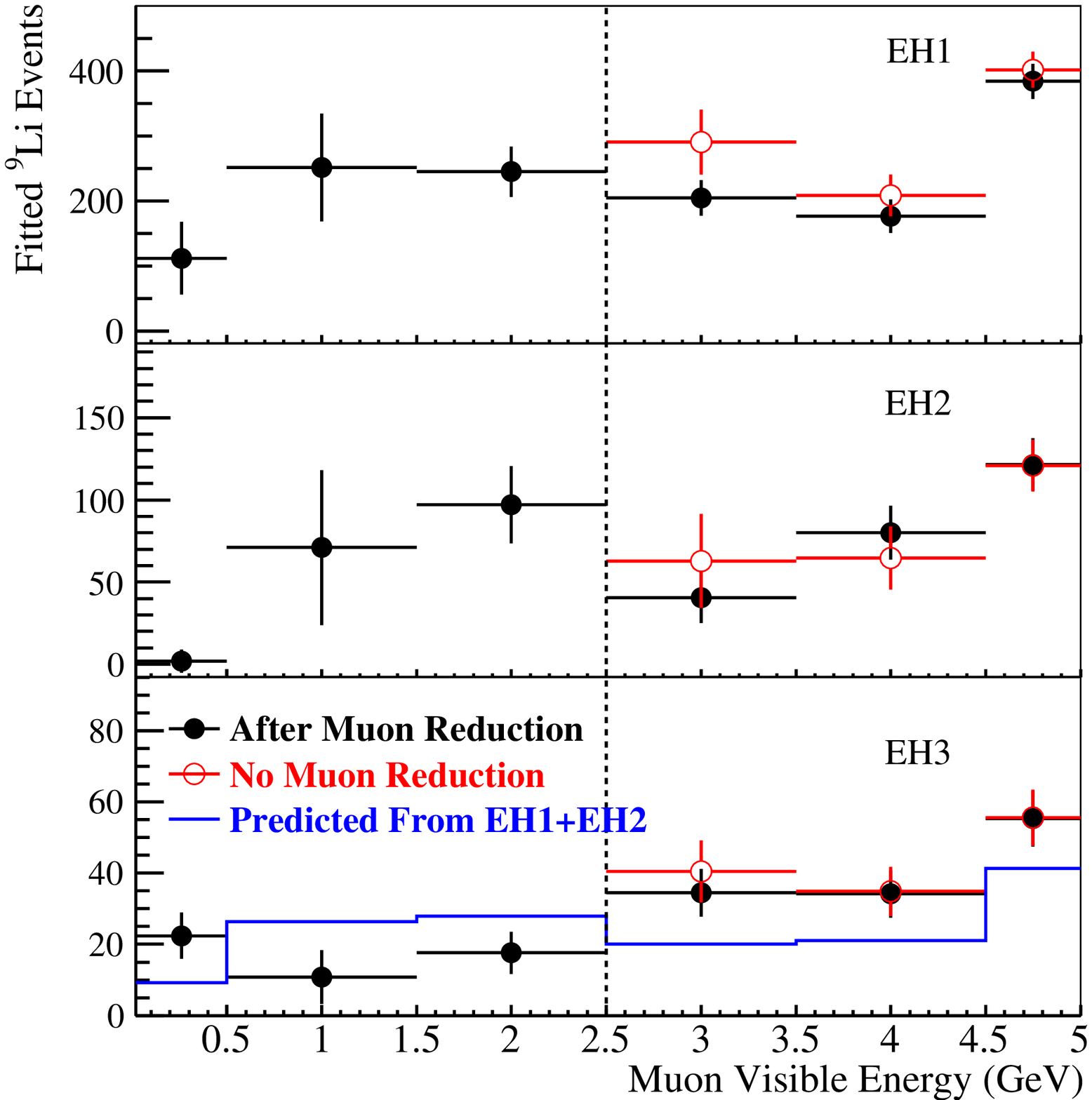}
\figcaption{The fitted $^9$Li yield as a function of the visible energy of parent muons for three experimental halls. The open circles represent the fit with all muons included. Due to high muon rate, the fit is done only for $E_\mu>2.5$ GeV. The filled circles  are the results obtaining by requiring a neutron following the muon as described in the text. In the bottom panel the prediction from the near site measurements is shown as solid line.
\label{fig:s5_li3site}}
\end{center}
%\end{figure}

\par
The best-fit results are shown in Fig.~\ref{fig:s5_li3site}. Since the statistics were quite low in EH3, we also predicted the $^9$Li yield in EH3 from the EH1 and EH2 yields by assuming that the $^9$Li yield with the same visible energy at different sites were identical, as shown in the bottom panel in Fig.~\ref{fig:s5_li3site}. The measured values agreed with the prediction within statistics. Another check was done by predicting the EH3 $^9$Li yield assuming that it follows an $E^{0.74}_\mu$ power law, where $E_\mu$ is the simulated average muon energy (See Table~\ref{tab:murate}), and normalizing to EH1 and EH2 measurements. Again the fitted EH3 $^9$Li yield agree with the prediction within statistics. By considering binning effects, differences between the results with and without muon reduction, and the difference between the predicted EH3 result and the measured result, we assigned a 50\% systematic uncertainty to the final result.

\subsection{$(\alpha,\,\,n)$ Backgrounds}

The $^{13}$C(${\alpha, \,\, n}$)$^{16}$O background was determined by measuring alpha-decay rates in situ and then using the MC to calculate the neutron yield.  We identified four sources of alpha decays, the $^{238}$U, $^{232}$Th, $^{227}$Ac decay chains and $^{210}$Po.  The decay chains are $\beta$-$\alpha$ cascades with half lives of 164.3 $\mu$s, 0.3 $\mu$s, and 1.781 ms, respectively. Fig.~\ref{fig:s5_alphat} displays the correlation of the prompt-delayed energy distributions for various time intervals corresponding to these cascade decays: 1-3 $\mu$s at upper left (group A are $^{212}$Bi-$^{212}$Po decays from the $^{232}$Th decay chain), 10-160 $\mu$s at upper right (group B are IBD events where the neutron captures on hydrogen. Group C are $^{214}$Bi-$^{214}$Po decays from the $^{238}$U decay chain, and group D are $^{219}$Rn-$^{215}$Po decays from the $^{227}$Ac decay chain). In the 1-2 ms region at lower left, only group D and some accidental coincidence events remain.

%\begin{figure}[htb]
\begin{center}
\includegraphics[width=\columnwidth]{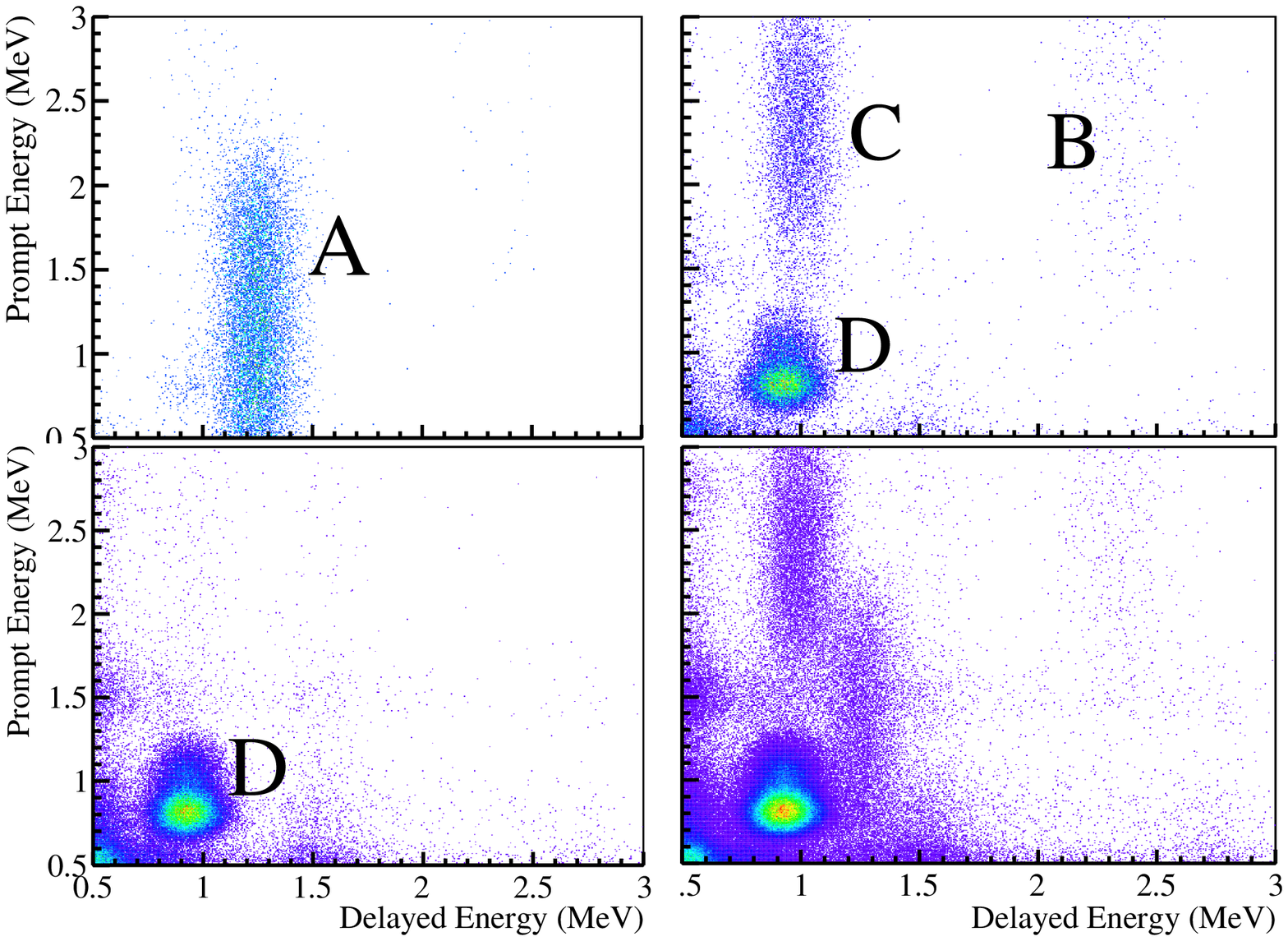}
\figcaption{Correlations of prompt and delayed energy for cascade decay chains of contaminants within the ADs.  At upper left are events with a time correlation between 1 and 3 $\mu$s, 10 to 160 $\mu$s is at upper right, 1-2 ms is at lower left, with the combined distributions at lower right.
\label{fig:s5_alphat}}
\end{center}
%\end{figure}

\par
$^{210}$Po was produced by the decay of $^{222}$Rn.  Its 5.3~MeV alpha produced 0.5~MeV of visible energy in an AD. The spatial distribution suggests that the $^{210}$Po background was due in part to an accumulation on the wall of the inner acrylic vessel.

\par
Geant4 was used to model the energy deposition process.  Based on the ($\alpha$, n) cross sections archived in JENDL~\cite{jendl}, the neutron yield as a function of $\alpha$ energy was calculated and summed. Finally, with the in-situ measured alpha-decay rates and MC determined neutron yields, the $^{13}$C(${\alpha, \,\, n}$)$^{16}$O rate was calculated, as listed in Table~\ref{tab:ibd}. The uncertainties come from the selection efficiencies of the Bi-Po and Rn-Po chain measurements, the possible deviation from equilibrium of the $^{238}$U, $^{232}$Th, and $^{227}$Ac decay chains, the fitting to determine the $^{210}$Po activity, and the simulation of ($\alpha$, n) reactions. During the Gd-LS synthesis, $^{238}$U, $^{232}$Th, and Ra were removed by radio-purification. They may contribute $\sim$30\% of the alphas of the whole chain. Thus a 30\% uncertainty was assigned for the possible deviation from equilibrium, which was the largest component in the uncertainties. A 10\% uncertainty was assigned to the neutron yield by comparing the MC simulation with an analytical calculation. Together with the other two components, $\sim$50\% uncertainties were estimated for the ($\alpha$, n) backgrounds, slightly different for each AD due to different alpha components in them.

\subsection{Correlated Backgrounds from Am-C Source}

\par
During data taking, the Am-C sources sat inside the ACUs on top of each AD.  Neutrons emitted from these sources would occasionally mimic IBD events by scattering inelastically with nuclei in the shielding material (emitting gamma rays) before being captured on a metal nuclei, such as Fe, Cr, Mn or Ni (releasing more gamma rays).  It was possible for the gamma-rays from both processes to enter the scintillating region and satisfy the IBD selection requirements.  Fig.~\ref{fig:s5_amc_peak} shows the energy spectrum in the three ADs at the far site of these delayed candidates from the Am-C sources. The rate in MC was normalized to data. There is good agreement between the data and MC.

%\begin{figure}[htb]
\begin{center}
\includegraphics[width=\columnwidth]{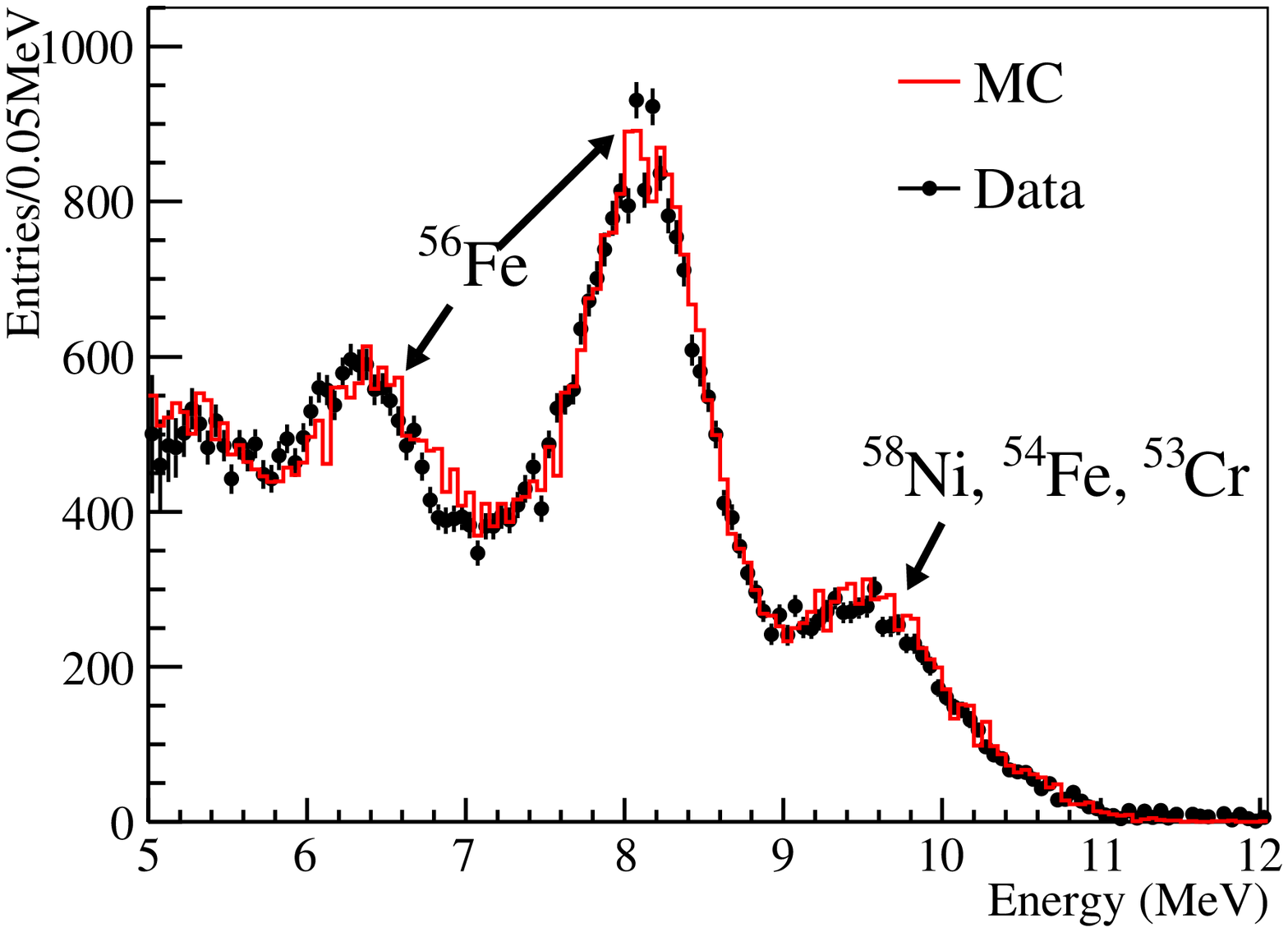}
\figcaption{Energy spectrum for events near the top of the three ADs in the far hall show three peaks consistent with neutron capture on $^{56}$Fe and $^{58}$Ni/$^{54}$Fe/$^{53}$Cr.
\label{fig:s5_amc_peak}}
\end{center}
%\end{figure}

\par
Fig.~\ref{fig:s5_amcz} shows an asymmetry of delayed-type events along the z axis as was seen by ADs in the far hall.  We estimated the delayed-type events from the Am-C sources by subtracting the number of delayed-type singles in the Z $<$ 0 region from the Z $>$ 0 region.  The Am-C correlated background rate was estimated by MC simulation normalized with the Am-C delayed-type event rate obtained from the data,
\begin{equation}\label{eqn:Am-C bg}
R_{corr} = R_{n-like\, data} \frac{N_{corr-MC}}{N_{n-like\, MC}}
\end{equation}
\noindent where $N_{corr-MC}$ and $N_{n-like\, MC}$ are the number of correlated background and number of delayed-type events in MC respectively, and $R_{n-like\, data}$ is the Am-C delayed-type event rate from data.  Even though the agreement in shape between data and MC is excellent for Am-C delayed-type events, we assigned 100\% uncertainty to the estimated background due to the Am-C sources to account for any potential uncertainty in the neutron scattering/capture cross sections used in the simulation.

%\begin{figure}[htb]
\begin{center}
\includegraphics[width=\columnwidth]{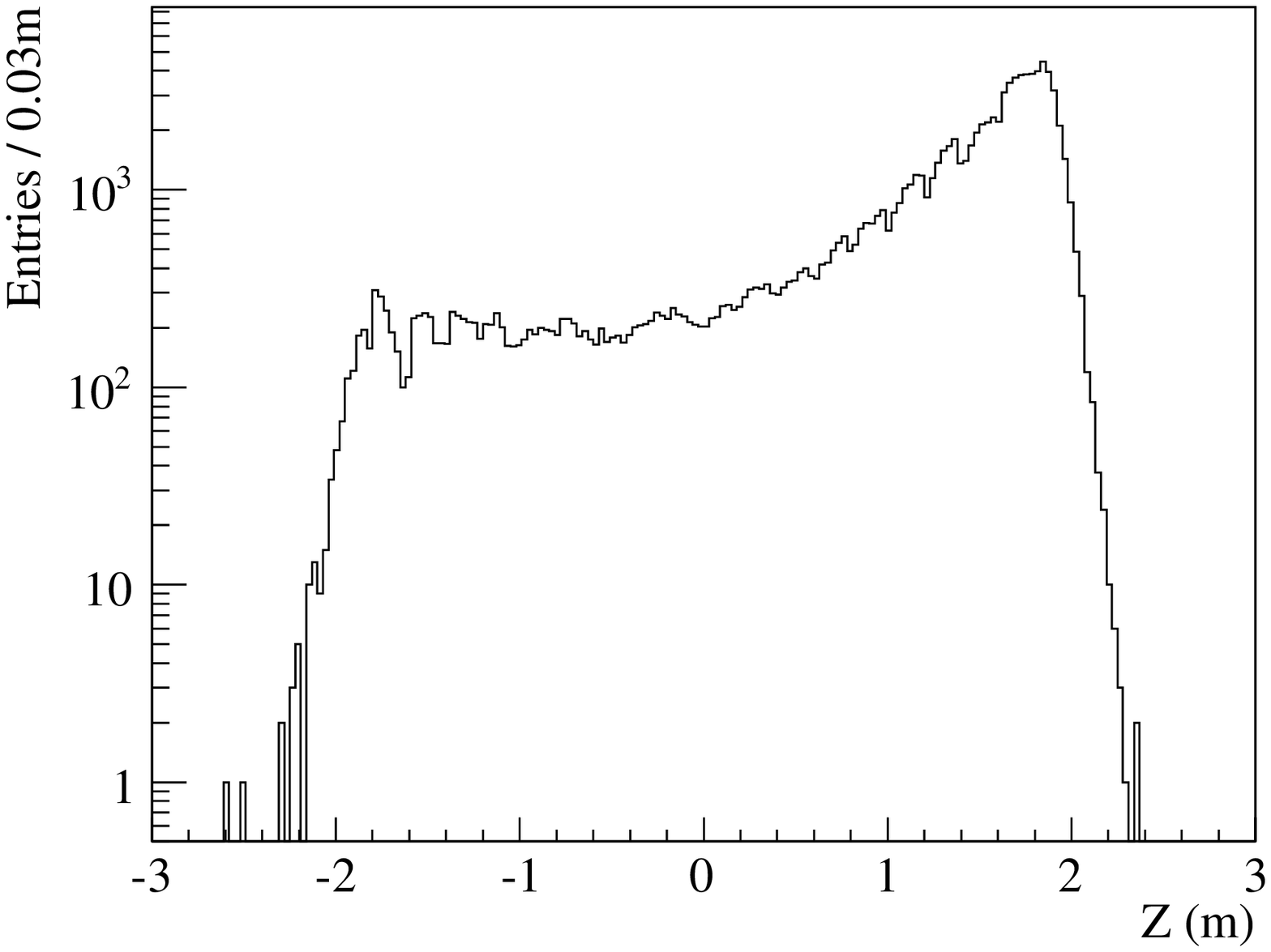}
\figcaption{Z distribution of delayed-type events. The excess in the top half of the ADs (Z $> 0$) comes from the Am-C sources in the ACUs. \label{fig:s5_amcz}}
\end{center}
%\end{figure}

\section{Side-by-Side Comparison in EH1}

\par
Relative uncertainties were studied with data by comparing two side-by-side antineutrino detectors. A detailed comparison using three months of data from the ADs in EH1 has been presented elsewhere~\cite{ad12}. An updated comparison of the prompt energy spectra of IBD events for the ADs in EH1 using 231 days of data (Sep.~23, 2011 to May 11, 2012) is shown in Fig.~\ref{fig:s7_AD12ratio} after correcting for efficiencies and subtracting background.  A bin-by-bin ratio of the AD1 and AD2 spectra is also shown. The ratio of the total IBD rates in AD1 and AD2 was measured to be $0.987\pm 0.004({\rm stat.}) \pm 0.003({\rm syst.})$, consistent with the expected ratio of 0.982.  The deviation of the ratio from unity was primarily due to differences in the baselines of the two ADs with a slight dependence on the individual reactor on/off status. It was shown that AD2 has a 0.3\% lower energy response than AD1 for uniformly distributed events, resulting in a slight tilt to the distribution shown in the bottom panel of Fig.~\ref{fig:s7_AD12ratio}.  The distribution of the data points denoted by open circles was created by scaling the AD2 energy by 0.3\%.  The bin-by-bin ratio with scaled AD2 energy agrees well with a flat distribution.

%\begin{figure}[htb]
\begin{center}
\includegraphics[width=\columnwidth]{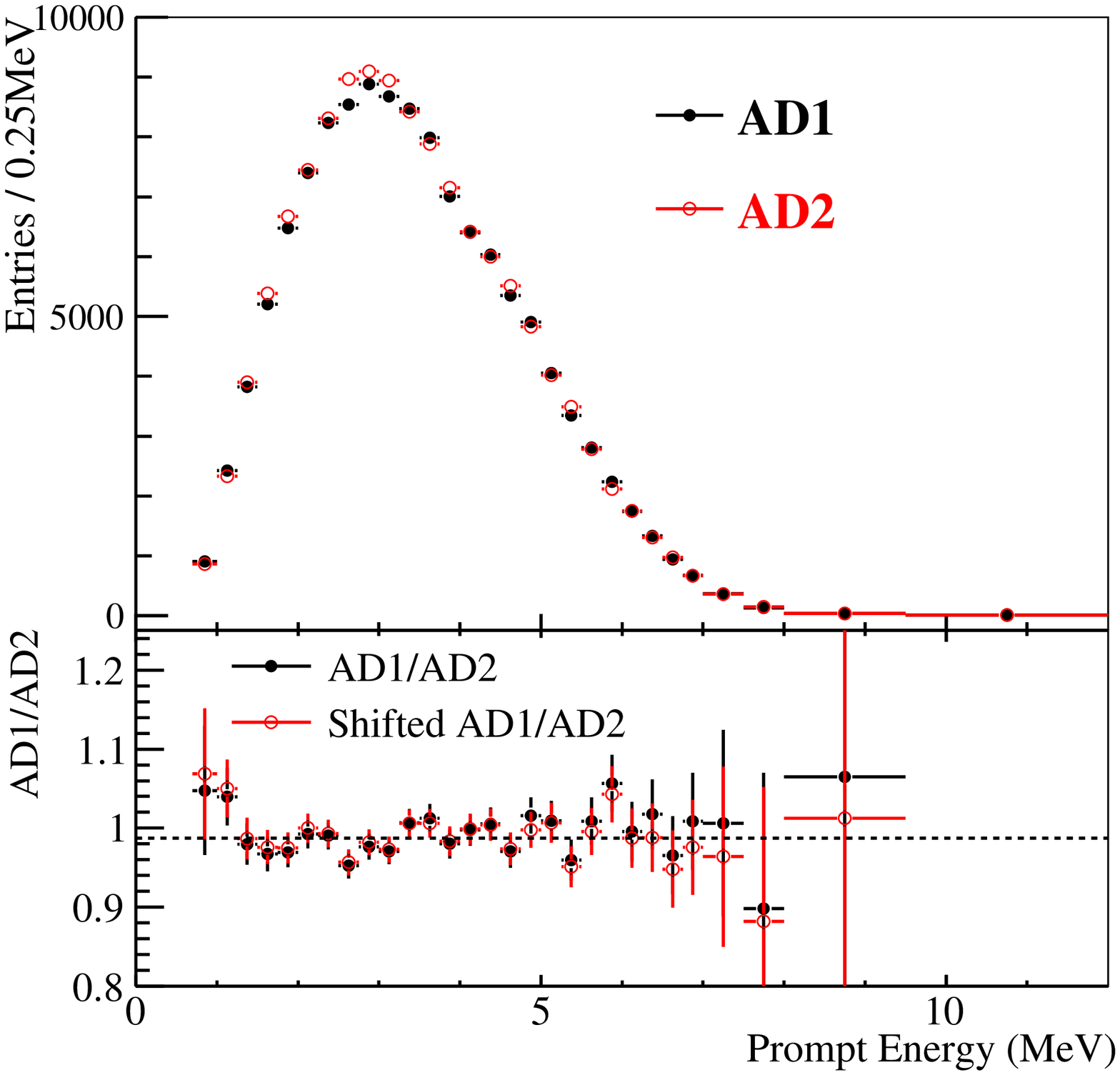}
\figcaption{
The energy spectra for the prompt signal of IBD events in AD1 and AD2 are shown in the top panel, along with the bin-by-bin ratio in the bottom panel (solid circles). In the bottom panel, the dashed line represents the ratio of the total rates for the two ADs, and the open circles show the ratio with the AD2 energy scaled by +0.3\%.
\label{fig:s7_AD12ratio}}
\end{center}
%\end{figure}

\section{Reactor Antineutrino Flux}

\par
Reactor antineutrinos result primarily from the beta decay of the fission products of four main isotopes, $^{235}$U, $^{239}$Pu, $^{238}$U, and $^{241}$Pu. The $\bar\nu_e$ flux of each reactor ($S(E)$) was predicted from the simulated fission rate ($F_i$) and the antineutrino spectrum per fission ($S_i$)~\cite{illschr,illvonf,illhahn,vogel238,mueller,huber} of each isotope~\cite{cjflux},
\begin{equation}
S(E)=\sum_i F_i S_i(E)
\end{equation}
where $i$ sums over the four isotopes. The fission rate was determined from the fission fraction $f_i$, the energy released per fission $E_i$, and the measured thermal power ($W_{\rm th}$),
\begin{equation}
F_i=\frac{W_{\rm th}f_i}{\sum_k f_k E_k}
\end{equation}
where both $i$ and $k$ are indices over the four isotopes.

\par
The thermal power data were provided by the power plant. The uncertainties were dominated by the flow rate measurements of feedwater through three parallel cooling loops in each core~\cite{cjflux,kme,helishi}. The correlations between the flow meters were not clearly known.  We conservatively assumed that they were correlated for a given core but uncorrelated between cores, giving a maximal uncertainty for the experiment. The assigned core-to-core uncorrelated uncertainty for thermal power was 0.5\%.

\par
A simulation of the reactor cores using commercial software (SCIENCE~\cite{sciencecode,appollo}) provided the fission fraction as a function of burn-up.  One example of fuel evolution is shown in Fig.~\ref{fig:s6_fission}. The fission fraction carries a 5\% uncertainty set by the validation of the simulation software. A complementary core simulation package was developed based on DRAGON~\cite{dragon} as a cross check and for systematic studies. The code was validated with the Takahama-3 benchmark~\cite{takahama} and agrees with the fission fraction provided by the power plant to 3\%.  Correlations among the four isotopes were studied using the DRAGON-based simulation package, and agrees well with the data collected in Ref.~\cite{djurcic}. Given the constraints of the thermal power and correlations, the uncertainties of the fission fraction simulation translated into a 0.6\% core-to-core uncorrelated uncertainty in the neutrino flux.

%\begin{figure}[htb]
\begin{center}
\includegraphics[width=\columnwidth]{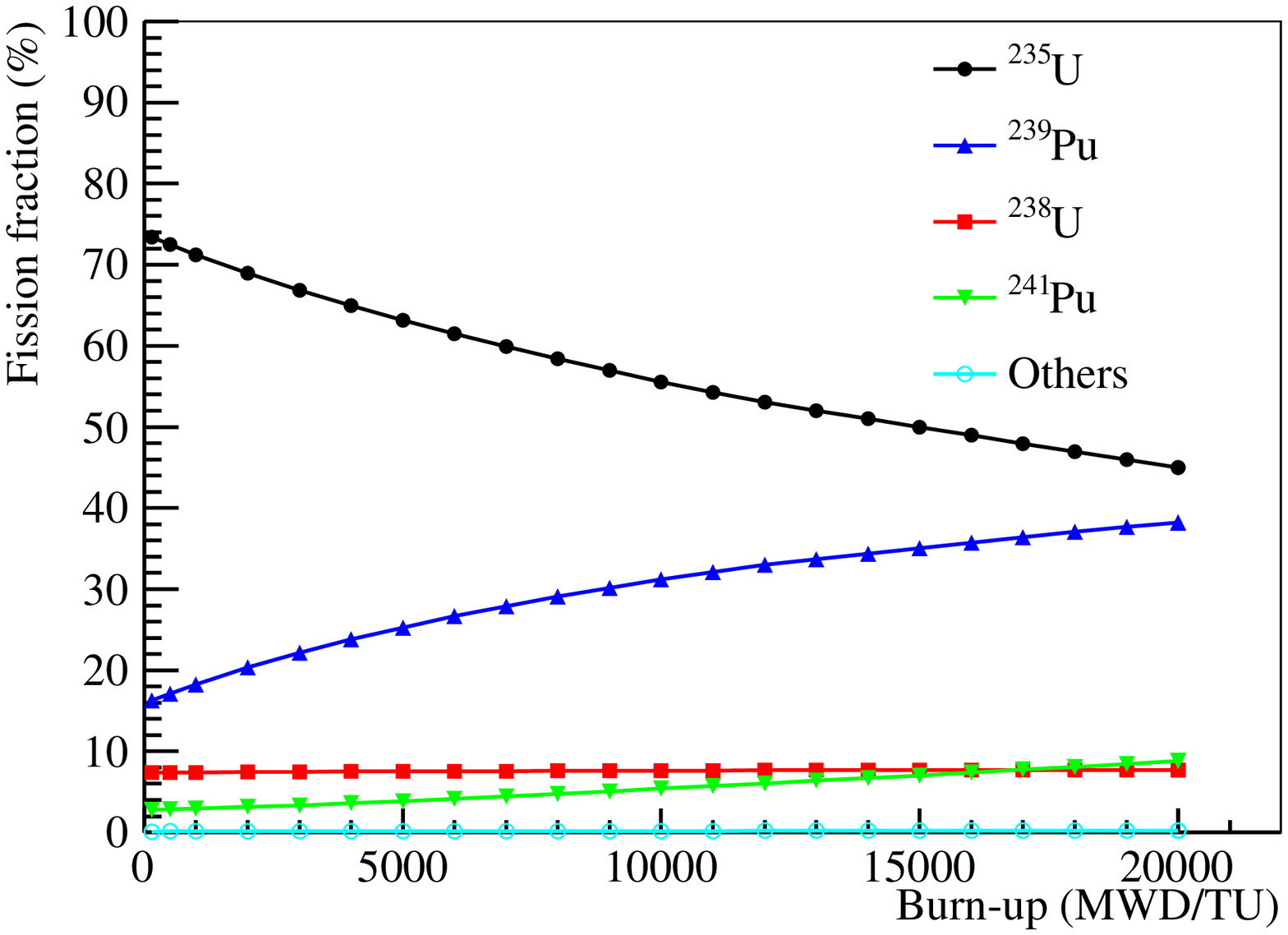}
\figcaption{
Fission fractions of reactor fuel isotopes as a function of burn-up from a simulation of reactor core D1.
Other isotopes contribute less than $0.3\%$ in total.
\label{fig:s6_fission}}
\end{center}
%\end{figure}

\par
The antineutrino spectra per fission is a correlated uncertainty that cancels out for a relative measurement. The reaction cross section for isotope $i$ was defined as $\sigma_{i}=\int S_i(E_\nu)\sigma(E_\nu)dE_\nu$, where $S_i(E_\nu)$ is the antineutrino spectra per fission and $\sigma(E_\nu)$ is the IBD cross section. We initially took the reaction cross section from Ref.~\cite{declais} but substituted the IBD cross section with that in Ref.~\cite{vogel}. The energy released per fission and its uncertainties were taken from Ref.~\cite{kopeikin}. Non-equilibrium corrections for long-lived isotopes were applied following Ref.~\cite{mueller}. Contributions from spent fuel~\cite{anfp,zhoub} ($\sim$0.3\%) were included as an uncertainty.

%\begin{table}[!htb]
\begin{center}
\tabcaption{Reactor-related uncertainties. \label{tab:reactoreff}}
\footnotesize
\begin{tabular}{lrrr}
\toprule
\hline
 \multicolumn{2}{c}{Correlated} & \multicolumn{2}{c}{Uncorrelated} \\
\hline
Energy/fission & 0.2\%  & Power  & 0.5\% \\
IBD reaction/fission & 3\%  & Fission fraction & 0.6\% \\
&& Spent fuel & 0.3\% \\
\hline
Combined & 3\%  & Combined  & 0.8\% \\
\bottomrule
\end{tabular}
\end{center}
%\end{table}

\par
The 3D spatial distribution of the isotopes within a core was also provided by the power plant. Simulation indicated that the spatial distribution has a negligible effect. As such, the reactor core was taken as a point source. The 3D core simulation with the input of the monthly in-core neutron flux measurement showed that the fission gravity center moves less than 1 cm on the horizontal plane and several cm vertically as the fuel burned. The resulting baseline variation can be ignored.

%\begin{figure}[htb]
\begin{center}
\includegraphics[width=\columnwidth]{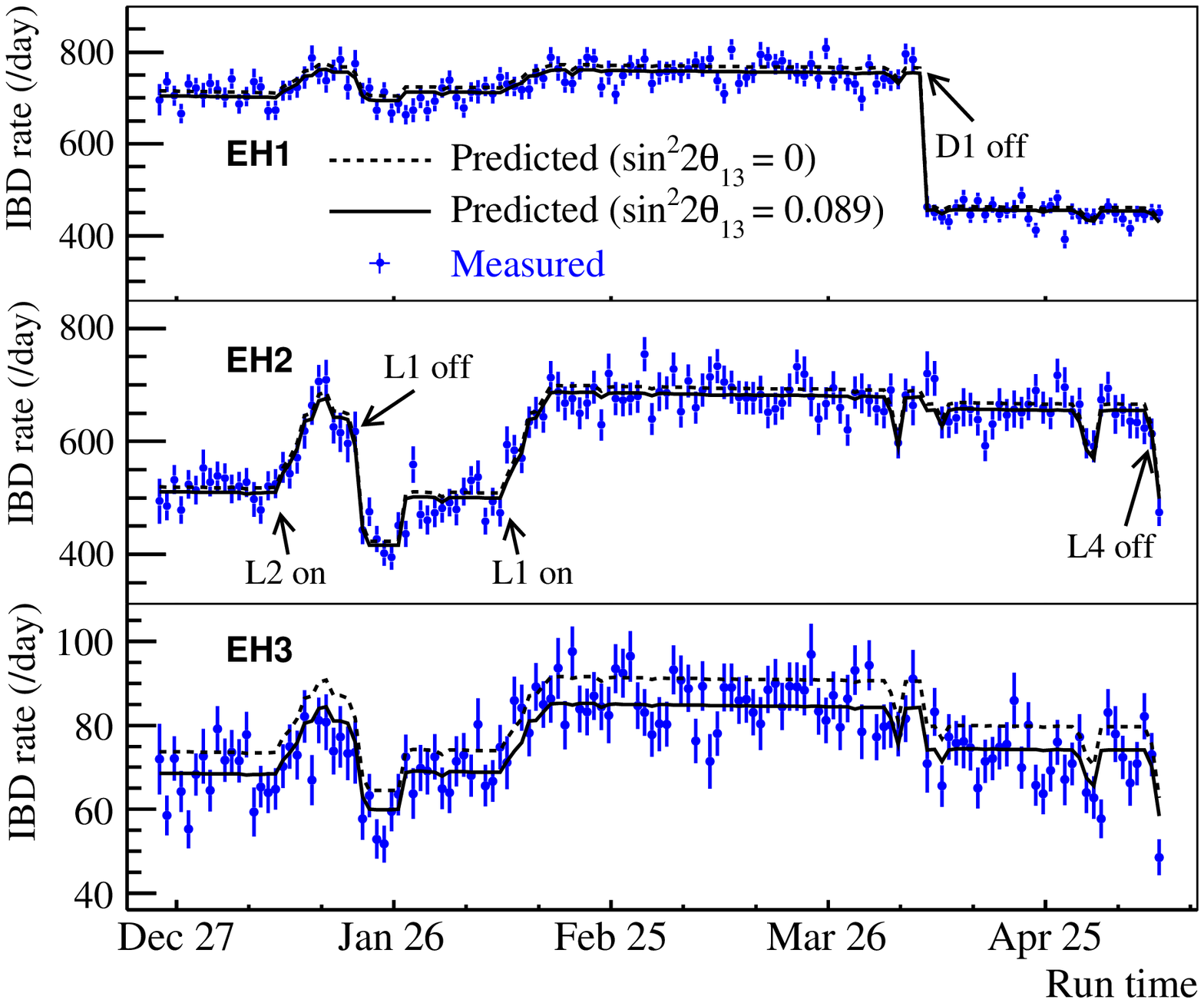}
\figcaption{
The measured daily average IBD rates per AD in the three experimental halls are shown as a function of time along with predictions based on reactor flux analyses and detector simulation. The reactor status is indicated.
\label{fig:s6_flux}}
\end{center}
%\end{figure}

\par
Fig.~\ref{fig:s6_flux} presents the background-subtracted and efficiency-corrected IBD rates in the three experimental halls. The predicted IBD rates from reactor flux calculation and detector simulation are shown for comparison. The dashed lines have been corrected with the best-fit normalization parameter $\varepsilon$ in Eq.~\ref{eqn:chi2} to reduce the biases from the absolute reactor flux uncertainty and the absolute detector efficiency uncertainty. These predictions are systematically higher than the data points due to the oscillation effect in the data at the near and far sites.  Predictions accounting for oscillation effects are also shown.

\section{Results}

\par
The \nuebar\ rate in the far hall was predicted with a weighted combination of the two near hall measurements assuming no oscillation. A ratio of the measured to expected rate is defined as
\begin{equation}
R=\frac{M_f}{\overline{N}_f}=\frac{M_f}{\alpha M_a + \beta M_b}\,,
\label{eqn:ratio}
\end{equation}
where $\overline{N}_f$ and $M_f$ are the predicted and measured rates in the far hall (sum of AD 4-6), $M_a$ and $M_b$ are the measured, background-subtracted IBD rates in EH1 (sum of AD 1-2) and EH2 (AD3), respectively. The weights $\alpha$ and $\beta$ are not unique since we approximate the contributions of the six reactors with the two observables $M_a$ and $M_b$. All valid physical models to determine the weights should satisfy the normalization requirement, i.e.\ the combination of the two near hall predictions should be equal to the direct prediction of the rate in the far hall in terms of antineutrinos emitted by the reactors. In this analysis we also required a maximum cancellation of reactor uncertainties, ignoring the statistical and detector-related systematic uncertainties. The values for $\alpha$ and $\beta$ were dominated by the baselines, and only slightly dependant on the integrated flux of each core. For the analyzed data set, $\alpha=0.0444$ and $\beta=0.2991$. The residual reactor-related uncertainty in $R$ was 5\% of the uncorrelated uncertainty of a single core. The ratio observed at the far hall was:
\begin{equation}
R=0.944\pm 0.007({\rm stat.}) \pm 0.003({\rm syst.})\,,\nonumber
\end{equation}
where the statistical (systematic) uncertainties were obtained by propagating statistical (uncorrelated systematic) uncertainties in the measured IBD counts in the three halls.

\par
The value of $\sin^22\theta_{13}$ was determined with a $\chi^2$ constructed with pull terms accounting for the correlation of the systematic errors~\cite{stump},
\begin{eqnarray}  \label{eqn:chi2}
 \chi^2 &=&
 \sum_{d=1}^{6}
 \frac{\left[M_d-T_d\left(1+  \varepsilon
 + \sum_r\omega_r^d\alpha_r
 + \varepsilon_d\right) +\eta_d\right]^2}
 {M_d+B_d}  \nonumber \\
 &+&
 \sum_r\frac{\alpha_r^2}{\sigma_r^2}
 + \sum_{d=1}^{6} \left(
 \frac{\varepsilon_d^2}{\sigma_d^2}
 + \frac{\eta_d^2}{\sigma_{B}^2}
 \right)
 \,,
\end{eqnarray}
where $M_d$ are the measured IBD events of the $d$-th AD with its backgrounds subtracted, $B_d$ is the corresponding background, $T_d$ is the prediction from antineutrino flux, including MC corrections and neutrino oscillations, $\omega_r^d$ is the fraction of IBD contribution of the $r$-th reactor to the $d$-th AD determined by the baselines and antineutrino fluxes. The uncorrelated reactor uncertainty is $\sigma_r$ (0.8\%), as shown in Table~\ref{tab:reactoreff}. The parameter $\sigma_d$ (0.2\%) is the uncorrelated detection uncertainty, listed in Table~\ref{tab:eff}. The parameter $\sigma_{B}$ is the quadratic sum of the background uncertainties listed in Table~\ref{tab:ibd}. The corresponding pull parameters are ($\alpha_r, \varepsilon_d, \eta_d$). The detector- and reactor-related correlated uncertainties were not included in the analysis. The absolute normalization $\varepsilon$ was determined from the fit to the data.

\par
The survival probability used in the $\chi^2$ was
\begin{eqnarray}
P_{sur} & = & 1 - \sin^22\theta_{13}\sin^2(1.267\Delta m^2_{31}L/E) \nonumber \\
 & - & \cos^4\theta_{13}\sin^22\theta_{12}\sin^2(1.267\Delta m^2_{21}L/E)\,, \nonumber
\end{eqnarray}
where $ \Delta m^2_{31} = 2.32\times10^{-3} {\rm eV}^2,
\sin^22\theta_{12} = 0.861^{+0.026}_{-0.022}$, and
$ \Delta m^2_{21} = 7.59^{+0.20}_{-0.21}\times 10^{-5} {\rm eV}^2$~\cite{SNOPRC}. The uncertainty in $\Delta m_{31}^2$~\cite{minosdm} had negligible effect and thus was not included in the fit.

%\begin{figure}[htb]
\begin{center}
\includegraphics[width=\columnwidth]{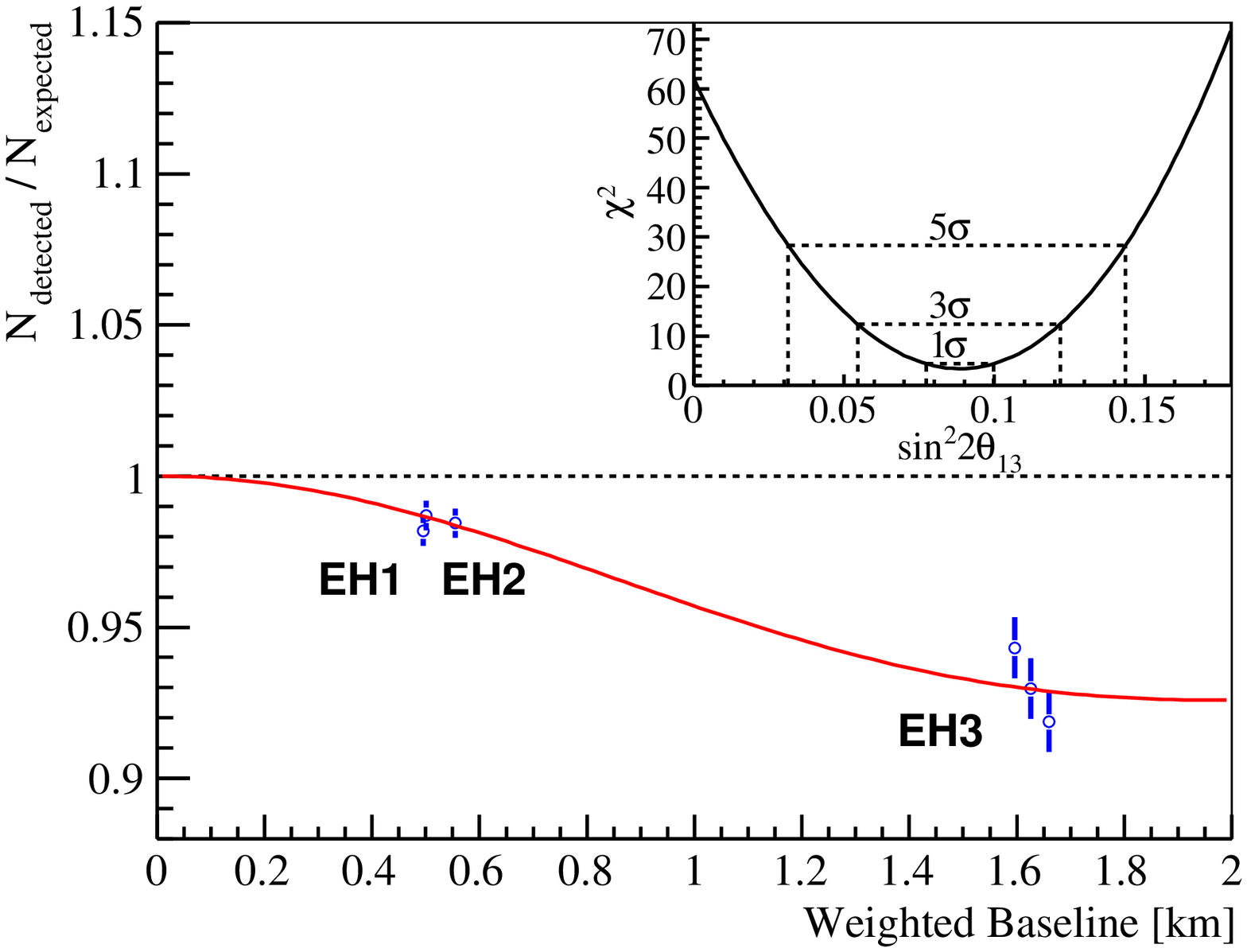}
\figcaption{
Ratio of measured versus expected signals in each detector, assuming no oscillation. The error bar is the uncorrelated uncertainty of each AD, including statistical, detector-related, and background-related uncertainties. The expected signal has been corrected with the best-fit normalization parameter. Reactor and survey data were used to compute the flux-weighted average baselines. The oscillation survival probability at the best-fit value is given by the smooth curve. The AD4 and AD6 data points were displaced by -30 and +30 m for visual clarity. The $\chi^2$ value versus $\sin^22\theta_{13}$ is shown in the inset. \label{fig:s7_chi2}}
\end{center}
%\end{figure}

\par
The best-fit value is
\begin{equation}
\sin^22\theta_{13}=0.089\pm 0.010({\rm stat.})\pm0.005({\rm syst.})
\nonumber
\end{equation}
with a $\chi^2$/NDF of 3.4/4. All best estimates of pull parameters are within its one standard deviation based on the corresponding systematic uncertainties. The no-oscillation hypothesis is excluded at 7.7 standard deviations. Fig.~\ref{fig:s7_chi2} shows the number of IBD candidates in each detector after correction for relative efficiency and background, relative to those expected assuming no oscillation. A $\sim$1.5\% oscillation effect appears in the near halls, largely due to oscillation of the antineutrinos from the reactor cores in the farther cluster. The oscillation survival probability at the best-fit values is given by the smooth curve.  The $\chi^2$ value versus sin$^22\theta_{13}$ is shown in the inset.

\par

The observed \nuebar\  spectrum in the far hall was compared to a prediction based on the near hall measurements $\alpha M_a + \beta M_b$ in Fig.~\ref{fig:s7_rate_deficit}. The distortion of the spectra is consistent with that expected due to oscillations at the best-fit $\theta_{13}$ obtained from the rate-based analysis.

%\begin{figure}[htb]
\begin{center}
\includegraphics[width=\columnwidth]{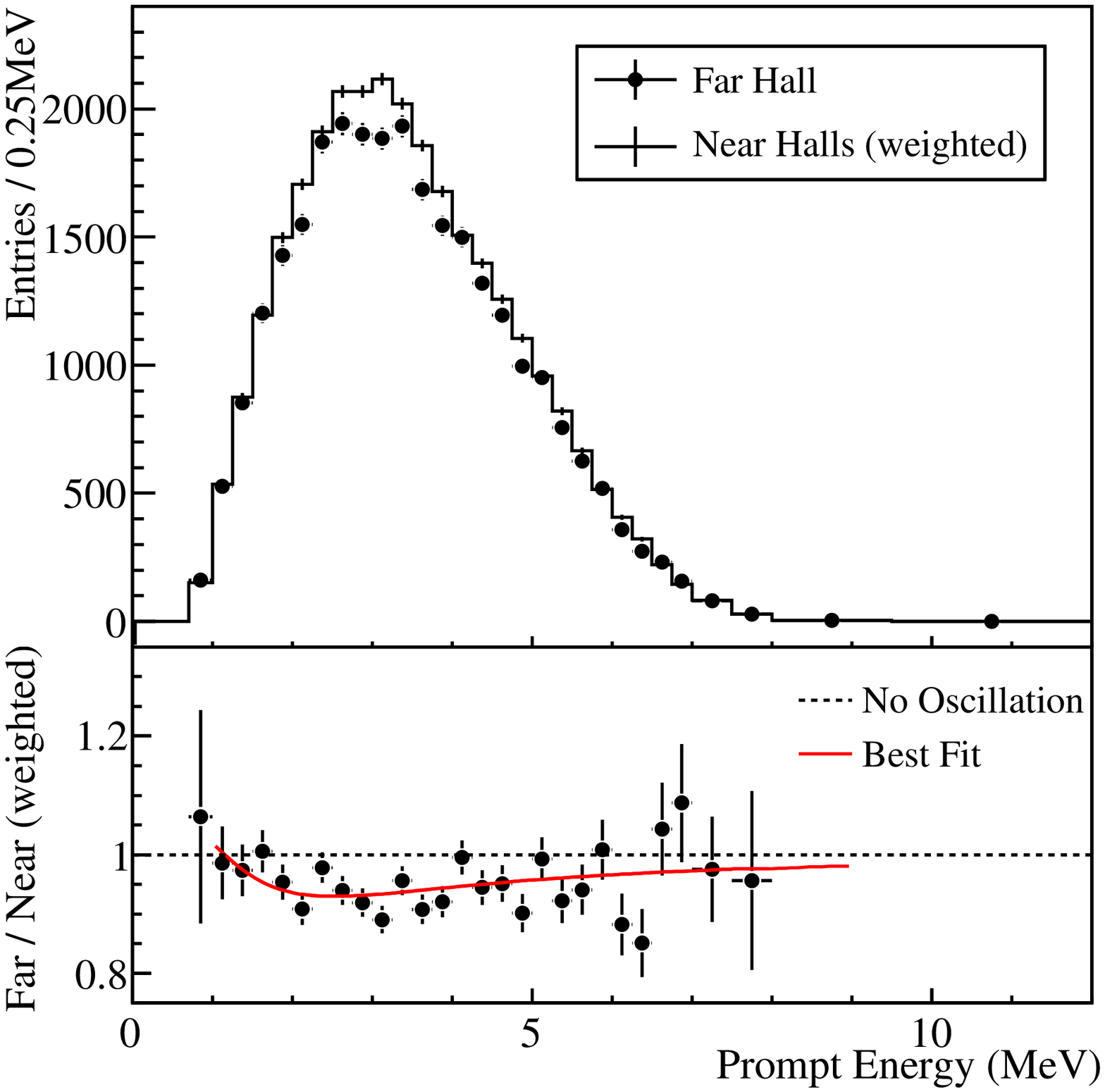}
\figcaption{Top: Measured prompt energy spectrum of the far hall (sum of three ADs)
compared with the no-oscillation prediction based on the measurements of the two near halls. Spectra were background subtracted. Uncertainties are statistical only. Bottom: The ratio of measured and predicted no-oscillation spectra. The solid curve is the expected ratio with oscillations, calculated as a function of neutrino energy assuming $\sin^22\theta_{13}=0.089$ obtained from the rate-based analysis. The dashed line is the no-oscillation prediction.  \label{fig:s7_rate_deficit}}
\end{center}
%\end{figure}

\section{Conclusions}
\par
We have updated the measurement of the neutrino mixing angle $\theta_{13}$ with a 116.8 kton-GW$_{\rm th}$-day livetime exposure at the far hall. A total of 138,835, 66,473, and 28,909 electron antineutrino candidates were detected in the Daya Bay near hall, the Ling Ao near hall, and the far hall, respectively. Compared with the prediction based on the near-hall measurements, a deficit of 5.6\% was observed in the far hall.  The rate-based analysis has yielded $\sin^22\theta_{13}=0.089\pm 0.010({\rm stat.})\pm0.005({\rm syst.})$.  This is the most precise measurement of $\sin^22\theta_{13}$ to date with a precision of 12.6\%, and supersedes our previous measurement~\cite{DB_discovery}. We anticipate additional improvements following the installation of two additional ADs in advance of an extended data run.

\section{Acknowledgement}

\par
The Daya Bay experiment is supported in part by the Ministry of Science and Technology of China, the United States Department of Energy,  the Chinese Academy of Sciences, the National Natural Science Foundation of China, the Guangdong provincial government, the Shenzhen municipal government,  the China Guangdong Nuclear Power Group, Shanghai Laboratory for Particle Physics and Cosmology, the Research Grants Council of the Hong Kong Special Administrative Region of China, University Development Fund of The University of Hong Kong, the MOE program for Research of Excellence at National Taiwan University, National Chiao-Tung University, and NSC fund support from Taiwan, the U.S. National Science Foundation, the Alfred~P.~Sloan Foundation, the Ministry of Education, Youth and Sports of the Czech Republic, the Czech Science Foundation, and the Joint Institute of Nuclear Research in Dubna, Russia. We thank Yellow River Engineering Consulting Co., Ltd.\ and China railway 15th Bureau Group Co., Ltd.\ for building the underground laboratory. We are grateful for the ongoing cooperation from the China Guangdong Nuclear Power Group and China Light~\&~Power Company.

\end{multicols}

\vspace{-1mm}
\centerline{\rule{80mm}{0.1pt}}
\vspace{2mm}
\begin{multicols}{2}

\end{multicols}

\end{document}

%% file: dyb_authors_cpc.tex
\newcommand {\IHEP} {1}
\newcommand {\USTC} {2}
\newcommand {\UW} {3}
\newcommand {\BNL} {4}
\newcommand {\NUU} {5}
\newcommand {\Caltech} {6}
\newcommand {\NCTU} {7}
\newcommand {\NJU} {8}
\newcommand {\Tsinghua} {9}
\newcommand {\CUHK} {10}
\newcommand {\SZU} {11}
\newcommand {\NCEPU} {12}
\newcommand {\Siena} {13}
\newcommand {\IIT} {14}
\newcommand {\LBNL} {15}
\newcommand {\UCBerkeley} {16}
\newcommand {\UIUC} {17}
\newcommand {\CDUT} {18}
\newcommand {\JINR} {19}
\newcommand {\SJTU} {20}
\newcommand {\BNU} {21}
\newcommand {\Princeton} {22}
\newcommand {\VT} {23}
\newcommand {\NTU} {24}
\newcommand {\CIAE} {25}
\newcommand {\UCLA} {26}
\newcommand {\SDU} {27}
\newcommand {\NKU} {28}
\newcommand {\Cincinnati} {29}
\newcommand {\DGUT} {30}
\newcommand {\HKU} {31}
\newcommand {\Houston} {32}
\newcommand {\Charles} {33}
\newcommand {\ZSU} {34}
\newcommand {\CWM} {35}
\newcommand {\RPI} {36}
\newcommand {\CGNPG} {37}
\newcommand {\Iowa} {38}
\newcommand {\XJTU} {39}

\author{
F.~P.~An$^{\IHEP}$ \and
Q.~An$^{\USTC}$ \and
J.~Z.~Bai$^{\IHEP}$ \and
A.~B.~Balantekin$^{\UW}$ \and
H.~R.~Band$^{\UW}$ \and
W.~Beriguete$^{\BNL}$ \and
M.~Bishai$^{\BNL}$ \and
S.~Blyth$^{\NUU}$ \and
R.~L.~Brown$^{\BNL}$ \and
G.~F.~Cao$^{\IHEP}$
% ...
}

\maketitle

\address{
\vspace{0.3cm}
{\normalsize (Daya Bay Collaboration)} \\
\vspace{0.3cm}
$^{\IHEP}$ (Institute~of~High~Energy~Physics, Beijing) \\
$^{\USTC}$ (University~of~Science~and~Technology~of~China, Hefei) \\
$^{\UW}$ (University~of~Wisconsin, Madison, WI) \\
$^{\BNL}$ (Brookhaven~National~Laboratory, Upton, NY) \\
$^{\NUU}$ (National~United~University, Miao-Li) \\
$^{\Caltech}$ (California~Institute~of~Technology, Pasadena, CA) \\
$^{\NCTU}$ (Institute~of~Physics, National~Chiao-Tung~University, Hsinchu) \\
$^{\NJU}$ (Nanjing~University, Nanjing) \\
$^{\Tsinghua}$ (Department of Engineering Physics, Tsinghua~University, Beijing) \\
$^{\CUHK}$ (Chinese~University~of~Hong~Kong, Hong~Kong) \\
$^{\SZU}$ (Shenzhen~Univeristy, Shen~Zhen) \\
$^{\NCEPU}$ (North China Electric Power University, Beijing)\\
$^{\Siena}$ (Siena~College, Loudonville, NY) \\
$^{\IIT}$ (Department of Physics, Illinois~Institute~of~Technology, Chicago, IL) \\
$^{\LBNL}$ (Lawrence~Berkeley~National~Laboratory, Berkeley, CA) \\
$^{\UCBerkeley}$ (Department of Physics, University~of~California, Berkeley, CA) \\
$^{\UIUC}$ (Department of Physics, University~of~Illinois~at~Urbana-Champaign, Urbana, IL) \\
$^{\CDUT}$ (Chengdu~University~of~Technology, Chengdu) \\
$^{\JINR}$ (Joint~Institute~for~Nuclear~Research, Dubna, Moscow~Region) \\
$^{\SJTU}$ (Shanghai~Jiao~Tong~University, Shanghai) \\
$^{\BNU}$ (Beijing~Normal~University, Beijing) \\
$^{\Princeton}$ (Joseph Henry Laboratories,~Princeton~University, Princeton, NJ) \\
$^{\VT}$ (Center~for~Neutrino~Physics, Virginia~Tech, Blacksburg, VA) \\
$^{\NTU}$ (Department~of~Physics, National~Taiwan~University, Taipei) \\
$^{\CIAE}$ (China~Institute~of~Atomic~Energy, Beijing) \\
$^{\UCLA}$ (University~of~California,~Los~Angeles, CA) \\
$^{\SDU}$ (Shandong~University, Jinan) \\
$^{\NKU}$ (School~of~Physics, Nankai~University, Tianjin) \\
$^{\Cincinnati}$ (University~of~Cincinnati, Cincinnati, OH) \\
$^{\DGUT}$ (Dongguan~University~of~Technology, Dongguan) \\
$^{\HKU}$ (Department~of~Physics, The University of Hong Kong, Pokfulam, Hong Kong) \\
$^{\Houston}$ (Department~of~Physics, University~of~Houston, Houston, TX) \\
$^{\Charles}$ (Charles University, Faculty of Mathematics and Physics, Prague) \\
$^{\ZSU}$ (Sun~Yat-Sen (Zhongshan)~University, Guangzhou) \\
$^{\CWM}$ (College~of~William~and~Mary, Williamsburg, VA) \\
$^{\RPI}$ (Rensselaer~Polytechnic~Institute, Troy, NY) \\
$^{\CGNPG}$ (China~Guangdong~Nuclear~Power~Group, Shenzhen) \\
$^{\Iowa}$ (Iowa~State~University, Ames, IA) \\
$^{\XJTU}$ (Xi'an~Jiaotong~University, Xi'an)\\
}